\documentclass[acmsmall,screen,nonacm]{acmart}

\usepackage{multirow}
\usepackage{array}
\usepackage{graphicx}
\usepackage{float}
\usepackage{tabularx}
\usepackage{boldline}
\usepackage{arydshln}
\usepackage{tablefootnote}
\usepackage{enumitem}
\usepackage{subfig}
\usepackage[ruled,vlined,linesnumbered]{algorithm2e}
\usepackage{setspace}
\usepackage{arydshln}
\usepackage{balance}
\usepackage{fontawesome}
\usepackage{url}
\usepackage[normalem]{ulem}

\newcommand*{\rvcolor}{\textcolor{black}} 

\usepackage[colorinlistoftodos,prependcaption,textsize=small]{todonotes}

\begin{document}

\title{Trajectory Similarity Measurement: An Efficiency Perspective}

\author{Yanchuan Chang}
\affiliation{%
  \institution{The University of Melbourne}
  \country{Australia}
}
\email{yanchuanc@student.unimelb.edu.au}

\author{Egemen Tanin}
\affiliation{\institution{The University of Melbourne}  \country{Australia}}
\email{etanin@unimelb.edu.au}

\author{Gao Cong}
\affiliation{\institution{Nanyang Technological University}  \country{Singapore}}
\email{gaocong@ntu.edu.sg}

\author{Christian S. Jensen}
\affiliation{\institution{Aalborg University}  \country{Denmark}}
\email{csj@cs.aau.dk}

\author{Jianzhong Qi}
\affiliation{\institution{The University of Melbourne}  \country{Australia}}
\email{jianzhong.qi@unimelb.edu.au}
\authornote{Corresponding author}

\begin{abstract}
Trajectories that capture object movement have numerous applications, in which similarity computation between trajectories often plays a key role. Traditionally, trajectory similarity is quantified by means of non-learned measures, e.g., Hausdorff, that operate directly on the trajectories. Recent studies  exploit deep learning to map trajectories to $d$-dimensional vectors, called embeddings. Then, some distance measure, e.g., Manhattan, is applied to the embeddings to quantify trajectory similarity. The resulting similarities are inaccurate: they only approximate the similarities obtained using the non-learned measures. As embedding distance computation is efficient, focus has been on obtaining embeddings of high accuracy.

Adopting an efficiency perspective, we analyze the time complexities of both the non-learned and the learning-based approaches, finding that the time complexities of the former approaches are not necessarily higher. Through extensive experiments on open datasets, we find that only a few learning-based approaches can deliver the promised higher efficiency, when the embeddings can be pre-computed, while non-learned approaches are more efficient for one-off computations. Among the learning-based approaches, the self-attention-based ones are the fastest and the most accurate. These results have implications for the use of trajectory similarity approaches given different application requirements.
Code is available at \url{https://github.com/changyanchuan/TrajSimiMeasures}.

\end{abstract}

\maketitle

\section{Introduction}\label{sec:introduction}
A trajectory is a sequence of timestamped point locations that captures the movement of an object such as a vehicle or a person.
The ability to quantify the similarity between two trajectories is essential in spatio-temporal data mining~\cite{representative_traj,e2dtc,dita,DFT,torch}. Due to the rich location and movement information encoded in trajectories and many application settings, no single universal trajectory similarity measure exists. Rather, different trajectory similarity measures have been proposed for different settings. These can be largely classified into two categories: \emph{non-learned measures} and \emph{learned measures}. 

\begin{figure}[thp]
    \centering
    \includegraphics[width=0.3\textwidth]{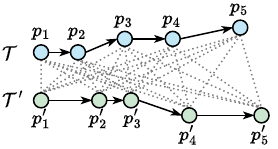}
    \caption{Computation of non-learned measures (dotted lines indicate point-to-point distance computation)}\label{fig:heuristic}
\end{figure}

Early studies focus on non-learned measures~\cite{edr,erp,edwp,hausdorff,frechet,dtw,lcss}. 
They typically work by matching the points between two trajectories (see Figure~\ref{fig:heuristic}) and are hand-crafted to capture similarity.
For example, \emph{Hausdorff}~\cite{hausdorff} computes a point matching that minimizes the sum of point-to-trajectory distances of two trajectories. 
Popular non-learned measures like the above have quadratic time complexity in the number of points on trajectories to examine. This complexity is considered a drawback in the literature~\cite{neutraj,t3s,trajcl,traj2simvec,cl-tsim,rsts}.

\begin{figure}[h]
  \centering
  \includegraphics[width=0.7\textwidth]{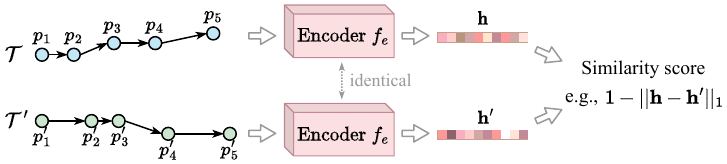}
  \caption{Computation of learned measures}\label{fig:learned}
\end{figure}

Learned measures~\cite{neutraj,t3s,trajgat,chen_iotj,tmn,gts,st2vec} mainly aim to improve the computational efficiency by exploiting deep learning and have recently attracted substantial interest. 
They generally follow the steps: (1) encode trajectories as vectors (called \emph{trajectory embeddings}), and (2) compute the vector distance (e.g., the Manhattan distance) between trajectory embeddings to serve as the trajectory distance (see Figure~\ref{fig:learned}). 
For example, \emph{t2vec}~\cite{t2vec}, \emph{NEUTRAJ}~\cite{neutraj}, and \emph{TMN}~\cite{tmn} use recurrent neural networks (RNNs)~\cite{lstm,gru}, while \emph{T3S}~\cite{t3s} and \emph{TrajCL}~\cite{trajcl} use self-attention models~\cite{transformer}.

Among learned measures, some (e.g., NEUTRAJ and T3S)  ``learn'' to \emph{approximate existing non-learned measures} (e.g., Hausdorff), i.e., the training signals are the ground-truth trajectory similarities provided by non-learned measures. Such sacrifice in accuracy is expected
to be rewarded by higher computational efficiency.

Another series of the learned measures (e.g., t2vec and TrajCL) improves not only the efficiency but also \emph{the measurement effectiveness}.
Such methods typically use self-supervised learning techniques to learn robust measures from unlabeled trajectories directly without relying on any non-learned measure. Once trained, they generally have better effectiveness, especially on measuring low-quality trajectories, than the non-learned ones.

A general perception of the learned measures in either category is that the learning-based approach yields superior efficiency, due to the simple vector format of embeddings and the highly optimized implementations of deep learning processes.

However, we observe that the time complexities of the learned measures are not necessarily lower than those of the non-learned ones. For example, T3S~\cite{t3s} and TrajCL~\cite{trajcl} also have quadratic time complexity in the number of trajectory points (covered in Section~\ref{subsec:related_learned}). Although previous studies~\cite{neutraj,t3s,trajcl,cl-tsim,tmn,rsts} report that the learned measures take less time to compute, they do not share detailed experimental settings, e.g., as to whether GPUs or CPUs are used for measuring trajectory similarity.

These observations prompt two questions:
\emph{(1)~Do learned measures have better computation time and space efficiencies than non-learned measures?
(2)~How do the learned measures compare with each other in terms of accuracy given training time constraints, e.g., due to different application requirements?}

This study provides the first answers to these questions based on comprehensive experiments on the efficiency of the trajectory similarity measures, as well as the accuracy of the learned measures given training time constraints.
This way, we aim to provide a foundation for the community to pursue promising research directions and to provide guidance on similarity measure selection.

We start by analyzing the time complexities of representative non-learned and learned measures and discuss their strengths and limitations (Section~\ref{sec:relatedwork}). We find that most of the existing learned measures do not necessarily have lower time complexity.

Then, we design three meta-algorithms for parallelizing the computation of non-learned measures (Appendix~\ref{sec:implementation}). We implement seven non-learned measures that leverage these algorithms to use CUDA streaming cores, thus enabling GPU-based empirical comparisons with learned measures. 
The meta-algorithms provide a comprehensive basis for future studies to implement new non-learned measures. 

We compare the efficiency of both types of measures using GPUs and CPUs on real datasets in a variety of settings (Section~\ref{sec:exp}). We use the measures for trajectory similarity computation, trajectory clustering~\cite{traclus,trajcluster_chenlu}, and trajectory $k$NN queries~\cite{DFT,trass,repose}. Previous studies~\cite{t2vec,neutraj,trajgat} use spatial indices like R-trees~\cite{r*tree} for trajectory $k$NN queries, which do not fit the learned measures. We use a $k$NN query framework designed for high-dimensional vectors to better realize the potential of the learned measures. 
We also investigate the impact of training time constraints on the accuracy of $k$NN queries for the learned measures, to guide similarity measure selection.

To sum up, we make the following contributions:
\begin{itemize}
    \item We review existing non-learned and learned spatio-temporal measures and analyze their time complexities, finding that the learned measures do not necessarily have the lowest  efficiency.
    
    \item We report on an extensive evaluation of the efficiency of both types of measures on GPUs and CPUs, finding that:
    \begin{enumerate}
    \item The non-learned measures are most efficient for one-off computation (online matching of incoming trajectories or $k$NN queries with data updates, e.g., for ride-sharing).
    \item The learned measures are most efficient for offline trajectory clustering and $k$NN queries, or when the trajectory embeddings can be pre-computed and reused, although they only offer approximate results.
    \item Further, among the learned measures, the self-attention-based ones are the fastest to train and offer the highest accuracy for $k$NN queries.
    \end{enumerate}
    
    \item We cover a simulated experiment to show a  learned measure that outperforms non-learned measures at one-off computation, and we provide future directions for achieving such a learned measure.
    
\end{itemize}

Several survey papers~\cite{hansu_survey,matt_survey,shengwang_survey} cover non-learned measures, one of which~\cite{shengwang_survey} also mentions a few early studies of learned  measures. None of these papers include a comprehensive comparison between the two types of  measures, which is our focus.


\section{Preliminaries}\label{sec:preliminaries}
We start by the core concepts.  Table~\ref{tab:symbol} lists the frequent symbols.

\begin{table}[htp]
\centering
\caption{Frequently used symbols}\label{tab:symbol}
\begin{tabular}{c|l}
\hlineB{3}
\textbf{Symbol} & \textbf{Description} \\ \hline \hline
$\mathcal{T}$ & A trajectory \\ \hline 
$p_i$ & The $i$-th point in a trajectory \\ \hline 
$n$ & The number of points in a trajectory \\ \hline
$\mathbf{h}$ & A trajectory embedding \\ \hline
$d$ & The dimensionality of trajectory embeddings \\ \hline
\hlineB{3}
\end{tabular}
\end{table}

\textbf{Trajectory.} A spatio-temporal trajectory $\mathcal{T}=[p_1, p_2, \ldots, p_n]$
is a sequence of $n$ points, where point $p_i$ is either given by a pair of coordinates $(x_i, y_i)$ or a pair of timestamped coordinates $(x_i, y_i, t_i)$. Trajectories without timestamps are also called paths.

\textbf{Non-learned trajectory similarity measure.} Given a trajectory dataset $D$, the similarity between two trajectories is defined by a function $f$: $D \times D \rightarrow \mathbb{R}_{\geqslant 0}$. 

A non-learned measure $f_h$: $D \times D \rightarrow \mathbb{R}_{\geqslant 0}$, e.g., Hausdorff, is a handcrafted function to capture the similarity between two trajectories. 
Distance measures can easily be converted into similarity measures, and thus we also consider them as similarity measures.

\textbf{Learned trajectory similarity measure.} A learned measure $f_l$, e.g., T3S, involves a two-step process. First, an encoding function $f_e$: $D \rightarrow \mathbb{R}^d$, which is learned, is applied to map each trajectory into a $d$-dimensional embedding space. Second, the distance between the embeddings $\mathbf{h}$ and $\mathbf{h}'$ of trajectories $\mathcal{T}$ and $\mathcal{T}'$ is used to quantify the similarity between $\mathcal{T}$ and $\mathcal{T}'$, e.g., $f_l(\mathcal{T}, \mathcal{T}') = 1 - || f_e(\mathcal{T}) - f_e(\mathcal{T}') ||_1 = 1 - || \mathbf{h} - \mathbf{h}' ||_1$, using the Manhattan distance.

\textbf{Trajectory similarity query.} Given a trajectory dataset $D$, a query trajectory $\mathcal{T}_q$, a trajectory similarity measure $f$, and a positive integer $k$, a \emph{trajectory similarity query} returns a set $S \subset D$ with $|S|=k$ such that $\forall \mathcal{T} \in S$, $\mathcal{T}' \in D\setminus S$ $\;( f(\mathcal{T}_q, \mathcal{T}) \geqslant f(\mathcal{T}_q, \mathcal{T}') )$. This query is also called a \emph{trajectory $k$NN query}. 

\textbf{Trajectory clustering.}
Given a trajectory dataset $D$ and a trajectory similarity measure $f$, \emph{trajectory clustering} groups the trajectories in $D$ into subsets (i.e., clusters) based on their similarity.

Both non-learned and learned measures can be applied in trajectory similarity queries and clustering. A learned measure typically approximates some non-learned measures. 
The approximation error is referred to as the \emph{inaccuracy} of the learned measure. 


\section{Trajectory Similarity Measures}\label{sec:relatedwork}
\begin{figure}[ht]
  \centering
  \hspace*{-4mm}
  \includegraphics[width=0.72\columnwidth]{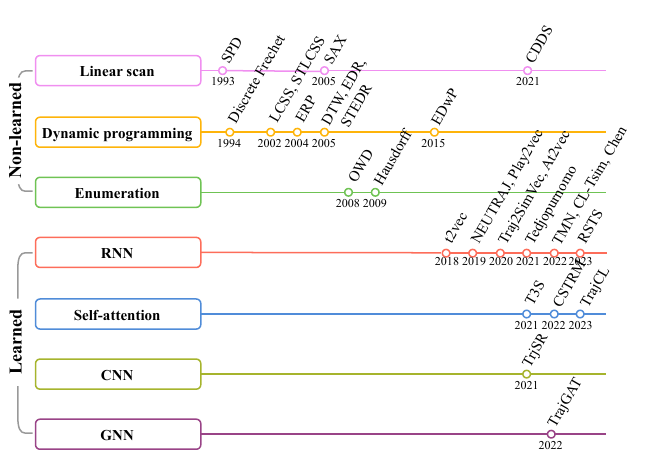}
  \caption{Representative trajectory similarity measures plotted in chronological order}\label{fig:tree}
\end{figure}

Next, we cover existing studies on spatial / spatio-temporal \emph{trajectory similarity measures}, including both \emph{non-learned} and \emph{learned} ones, and trajectory queries, including \emph{trajectory similarity queries}, and \emph{trajectory clustering}.
Figure~\ref{fig:tree} summarizes the representative measures based on Euclidean space that we focus on. As the figure shows, the learned measures dominate the recent literature.

\begin{table*}[t]
\centering
\caption{Categorization of representative trajectory similarity measures in Euclidean space}
\label{tab:model_summary}
\setlength{\tabcolsep}{1mm}
\renewcommand{\arraystretch}{1.2}
\resizebox{\textwidth}{!}{
\begin{tabular}{l|l|l|c|c}
\hlineB{3}
\textbf{Category} & \textbf{Methodology} & \textbf{Measure} & \textbf{Time Complexity} & \textbf{Space Complexity}\\ \hline \hline
\multirow{4}{*}{\textbf{Non-learned measures}} & Linear scan & \begin{tabular}[c]{@{}l@{}} SPD~\cite{euclidean_SPD} \end{tabular}, CDDS~\cite{stsjoin}, SAX~\cite{sax_traj1} & $O(n)$ & $\Theta(1)$ \\\cline{2-5} 
& Dynamic programming & \begin{tabular}[c]{@{}l@{}}  EDR~\cite{edr}, ERP~\cite{erp}, EDwP~\cite{edwp},  LCSS~\cite{lcss}, DTW~\cite{dtw}, \\ Discrete Fr\'echet~\cite{dfrechet}, STEDR~\cite{tedjopurnomo}, STLCSS~\cite{tedjopurnomo} \end{tabular}  & $O(n^2)$ & $\Theta(n)$ \\ \cline{2-5} 
& Enumeration & \begin{tabular}[c]{@{}l@{}}
 OWD~\cite{owd}, Hausdorff~\cite{hausdorff} \end{tabular} & $O(n^2)$ & $\Theta(1)$ \\ \hline \hline
\multirow{4}{*}{\textbf{Learned measures}} & Recurrent neural network & \begin{tabular}[c]{@{}l@{}}t2vec~\cite{t2vec}, NEUTRAJ~\cite{neutraj}, Play2vec~\cite{play2vec}, \\ At2vec~\cite{at2vec}, Traj2SimVec~\cite{traj2simvec}, Tedjopurnomo~\cite{tedjopurnomo}, \\ Chen~\cite{chen_iotj}, CL-Tsim~\cite{cl-tsim}, TMN~\cite{tmn}, RSTS~\cite{rsts}\end{tabular} & $\Omega(nd^2)$ & $\Omega(d^2)$\\ \cline{2-5} 
& Self-attention neural network & T3S~\cite{t3s}, CSTRM~\cite{cstrm}, TrajCL~\cite{trajcl} & $\Omega(n^2d)$ & $\Omega(d^2+nd+n^2)$ \\ \cline{2-5} 
& Convolutional neural network & TrjSR~\cite{trjsr} & $\Omega(m{k^2}{n_k}c)$ & $\Omega(k^2{n_k}c+mc)$ \\ \cline{2-5} 
& Graph neural network & TrajGAT~\cite{trajgat} & $\Omega(n{n_e}d)$ & $\Omega(d^2+nd+n{n_e})$ \\ 
\hlineB{3}
\end{tabular}
}
\end{table*}

\subsection{Non-learned Trajectory Similarity Measures}\label{subsec:related_heuristic}
Non-learned measures are generally based on point matching~\cite{stsjoin,edr,erp,frechet,hausdorff,edwp}. The similarity between two trajectories is derived from the distances between the matched point pairs. 
Based on how the point matches are computed, we categorize the  non-learned measures into three classes: (i)~\emph{linear scan-based}, (ii)~\emph{dynamic programming-based} and (iii)~\emph{enumeration-based}  measures.

\begin{figure}[htp]
    \centering
    \includegraphics[width=0.3\textwidth]{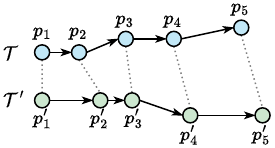}
    \caption{Computation of linear scan-based measures}\label{fig:linear}
\end{figure}

\textbf{Linear scan-based measures.} Linear scan-based measures~\cite{euclidean_SPD, stsjoin,sax_traj1} take only a single scan over two trajectories (cf.~Figure~\ref{fig:linear}, where each gray dotted line denotes a pair of matched points). Such measures take $O(n)$ time to compute, assuming $n$ points per trajectory, and $\Theta(1)$ space, to store the partial similarity results, excluding the $O(n)$ space to hold the trajectories (same below).  

For example,  
\emph{Sum-of-Pair Distance}  (SPD)~\cite{euclidean_SPD} simply  
matches the points on two trajectories following the order of the points and sums up the pairwise point distances.
It assumes trajectories of the same number of points. 
\emph{Close-Distance Duration Similarity} (CDDS)~\cite{stsjoin} further considers the time dimension. It scans the points on two trajectories and sums up the time span when the points on both trajectories are within a given spatial distance threshold. The resulting sum is used as the similarity. 
\emph{Symbolic Aggregate Approximation} (SAX) representation, which was designed for time series~\cite{sax}, can be extended to compute spatio-temporal trajectory similarity~\cite{sax_traj1}, as trajectories are multivariate time series. SAX divides a time series (a trajectory in our case, same below) into segments and uses a symbol to represent the average of the points within each segment. Two time series are aligned by the time dimension. The differences between each pair of aligned symbol values  are summed up as the distance between two time series. SAX essentially discretizes time series and computes their shape similarity.

\textbf{Dynamic programming-based measures.} The linear scan-based measures may compute sub-optimal point matches and hence result in falsely large trajectory similarity values. 
For example, SPD simply matches the points on two trajectories following the order that the points appear in the respective trajectories. It does not match points based on their spatial distances.
To address the issue, dynamic programming (DP)-based measures are proposed to explore a larger point-matching space while confining the computation costs. 
As Figure~\ref{fig:heuristic} shows, such measures examine all point pairs  incrementally with DP in $O(n^2)$ time, taking $\Theta(n)$ space to store the intermediate similarity (distance) values. 
 
For example, \emph{Dynamic Time Warping} (DTW)~\cite{dtw}, which was designed for time series analysis, e.g., speech recognition, has been extended to trajectories by using the Euclidean distance function, exploiting the property that it allows many-to-one point matching to cope with trajectories with different travel speeds or sampling rates. 
DTW computes a set of point alignments between two trajectories that achieves a global minimum sum of point-to-point distances. 
It  does not require  trajectories of the same length. \emph{Discrete Fr\'echet}~\cite{dfrechet} also allows 
many-to-one point matching. It returns the maximum distance between any matched point pairs.

\emph{Longest Common Sub-Sequence} (LCSS)~\cite{lcss} adapts a string similarity measure to reveal the longest common sub-sequence of two trajectories. Here, a common sub-sequence refers to consecutive pairs of points that are within a given spatial distance threshold. 
Another series of studies~\cite{edr,erp,edwp} adapt the edit distance which is also a string similarity measure. They compute the cost to ``edit'' (insert, delete, or substitute) points on a trajectory to match (i.e., be within a predefined distance threshold) those in the other trajectory. A large distance suggests that more edits are needed to create a match and hence less similar trajectories.
For example, \emph{Edit Distance on Real sequence} (EDR)~\cite{edr} considers a same unit cost for each edit. 
Next, \emph{Edit Distance with Real Penalty} (ERP)~\cite{erp}  factors the point distances into the edit costs. Further, \emph{Edit Distance with Projections} (EDwP)~\cite{edwp} uses an interpolation-style insertion operation. It adds points on the line between two adjacent points of a trajectory when point insertions are needed to form matches.
A few other measures consider both spatial and temporal distances, such as STEDR and STLCSS~\cite{tedjopurnomo}, which extend EDR and LCSS by adding a temporal distance threshold when matching the points, respectively.

\textbf{Enumeration-based measures.} These measures compute all pairwise point distances directly and aggregate them to form a trajectory similarity (cf.~Figure~\ref{fig:heuristic}). 
They take $O(n^2)$ time and $O(1)$ space, as they do not need to store intermediate results.

For example, the \emph{One Way Distance} (OWD)~\cite{owd} uses the average point-to-trajectory distance as the distance between two trajectories. The point-to-trajectory distance here is the minimum distance between a  point to any point on a trajectory.
\emph{Hausdorff}~\cite{hausdorff}, which was designed for image matching, computes the maximum point-to-trajectory distance. A small Hausdorff distance means each point in a trajectory to have a close match in another trajectory. Thus, the two trajectories form a close match.

\textbf{Discussion.} Some non-learned measures have approximate algorithms (e.g., \emph{Approx-DTW}~\cite{ying2016simple} and \emph{aprxFréchetI}~\cite{frechet_approx1}) with lower running times. The comparison between non-learned and learned approximations is not our focus and is left for future work.

\subsection{Learned Trajectory Similarity Measures}\label{subsec:related_learned}
Studies in the past five years focused on deep learning models to encode trajectories and subsequently learn trajectory similarity~\cite{t2vec,neutraj,t3s,trajgat,traj2simvec,tmn,trajcl,trjsr}. These studies can be 
categorized into two classes according to their design purposes: (i) to learn and approximate  
existing non-learned measures~\cite{neutraj,t3s,trajgat} with supervised learning, where some non-learned measure is used to provide the supervision signals, and (ii)~to learn latent similarity measures independent from any non-learned measures~\cite{t2vec,trjsr,trajcl}. For the latter class,  self-supervised learning is used to learn trajectory embeddings. 
The similarity between two trajectories are calculated based on their embeddings, e.g., using the $L_1$ distance.

A core component in these studies is the \emph{backbone trajectory encoder}, which takes a trajectory as input and outputs an embedding. The  encoders play a central role in the learned  measures. Thus, we review the learned measures based on the encoders used.

\begin{figure}[ht]
  \centering
  \hspace*{1.0cm}
  \includegraphics[width=0.45\textwidth]{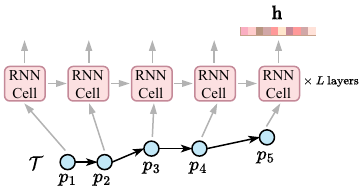}
  \caption{RNN-based learned measures}\label{fig:rnn}
\end{figure}

\textbf{Recurrent neural network (RNN)-based measures.} 
Since trajectories are sequences, RNNs form a natural backbone trajectory encoder.   
RNNs encode each trajectory point recurrently by considering both the historical states (i.e., aggregated information from preceding points) and the current state (i.e., the current point to be encoded). When an RNN terminates, the final output state entails aggregated information from all points on a trajectory, which is used as the trajectory embedding, i.e., $\mathbf{h}$ in Figure~\ref{fig:rnn}.

RNN-based measures (using Long Short-Term Memory, LSTM~\cite{lstm} or Gated Recurrent Unit, GRU~\cite{gru}) take at least $\Omega(nd^2)$ time to encode a trajectory because they need to compute the hidden representation of $n$ states (for the $n$ points on a trajectory), while each hidden representation takes $\Theta(d^2)$ time to compute (i.e., hidden feature mapping). Here, $d$ is the embedding dimensionality.  
Some approaches may exceed $\Theta(nd^2)$ time, e.g.,  $\Theta(nd^2 + n^2d)$ time for TMN~\cite{tmn}. RNN-based measures take $\Omega(d^2)$ space for the model parameters (i.e., $d \times d$ weight matrices) and another $\Omega(d)$ space to store the intermediate results during the encoding  process. 

Most RNN-based studies use a grid cell-based input representation. They partition the data space with a grid and transform a trajectory into a sequence of grid cells enclosing the trajectory. This transformation has two benefits: (1) The cell-based representation reduces the input space from points in a continuous space to a small number of discrete cells. This creates an input data distribution that is easier to be learned. (2) As a side effect, the cell-based representation alleviates the impact of GPS errors and varying  trajectory sampling rates, leading to more robust embeddings.

The first RNN-based model, \emph{t2vec}~\cite{t2vec}, adapts GRU by introducing a spatial proximity aware loss function, which penalizes the model when it generates large similarity prediction errors on spatially close trajectories. 
\emph{NEUTRAJ}~\cite{neutraj} uses GRU (according to its released code) and adds a spatial attention memory unit, such that the embedding learning for a trajectory can refer to spatially close trajectories seen by the GRU before. 
Further, NEUTRAJ has a weighted ranking loss function to encourage model  learning from the most similar trajectory pairs.  
\emph{Traj2SimVec}~\cite{traj2simvec} improves upon NEUTRAJ in two aspects:
(1)~Traj2SimVec leverages a $k$-d tree~\cite{kdtree} to select a set of most similar trajectories as the positive training samples, rather than randomly sampling as in NEUTRAJ. Hence, it achieves better training efficiency. 
(2)~Traj2SimVec uses a sub-trajectory-based loss to learn the detailed alignment and distances between points, while the loss function of NEUTRAJ is based on the embeddings of full trajectories. 
Chen et al.~\cite{chen_iotj} also build upon NEUTRAJ. They use an interpolation-based trajectory calibration process to generate smoother trajectories for model training. 
\emph{CL-Tsim}~\cite{cl-tsim} adopts contrastive learning to help generate more diverse training samples so as to obtain more robust embeddings.

Unlike the models above that learn to encode each trajectory separately, \emph{TMN}~\cite{tmn} introduces a dual-branch model to learn trajectory embeddings and point matches at the same time. 
Its matching module aims to simulate the computation process of the non-learned  measures. 
TMN requires to input two trajectories together. It cannot be used to encode individual trajectories. Thus, this model cannot be used for the $k$NN query experiments in  Section~\ref{subsec:exp:knn}.

Besides, Tedjopurnomo et al.~\cite{tedjopurnomo} and Li et al. (i.e., \emph{RSTS})~\cite{rsts}  adapt t2vec to measure spatio-temporal trajectory similarity. 
RSTS simply introduces a three-dimensional grid cell where the third dimension models the time. Tedjopurnomo et al. introduce three loss functions to jointly learn trajectory similarity at  trajectory, point, and pattern levels. 
Several studies adopt RNN models to learn similarity for special types of trajectories, e.g., point of interest (POI, \emph{At2vec})~\cite{at2vec} and sports play (\emph{Play2vec})~\cite{play2vec,play2vecplus} trajectories.

\begin{figure}[ht]
  \centering
  \hspace*{1.5cm}
  \includegraphics[width=0.5\textwidth]{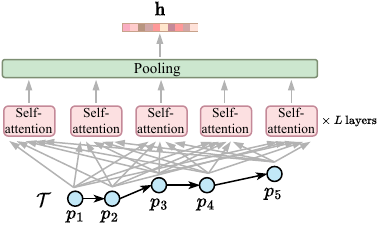}
  \caption{Self-attention-based learned measures}\label{fig:self-attention}
\end{figure}

\textbf{Self-attention-based measures.}
\emph{Multi-head Self-attention} (``self-attention'' in short)~\cite{transformer} is a more recent sequential model that addresses the \emph{catastrophic forgetting} issue of the RNNs. 
It learns the hidden correlation between every two elements in an input sequence (cf.~Figure~\ref{fig:self-attention}).
It takes $\Omega(n^2d)$ time to encode a trajectory. While this seems to be higher than the time taken by RNNs, self-attention models may run much faster than RNNs on GPU.  This is because self-attention models run in only one round to compute a sequence embedding, which is highly  parallelizable. In contrasts, RNNs need to run $n$ rounds iteratively due to their recurrent structures.
Self-attention-based measures take $\Omega(d^2+nd+n^2)$ space, where the model parameters (i.e., weight matrices) take $\Omega(d^2)$ space, and the intermediate results for computing the attention coefficients (between the $n^2$ pairs of points) take $\Omega(nd+n^2)$ space.

A few studies have used self-attention. 
\emph{CSTRM}~\cite{cstrm} leverages Masked Language Modeling~\cite{bert} to learn trajectory embeddings in a self-supervised manner. The trained trajectory encoder can be further fine-tuned to approximate non-learned measures. 
\emph{T3S}~\cite{t3s} combines self-attention and LSTM to capture the topological and spatial features of trajectories. 
Later, \emph{TrajCL}~\cite{trajcl} introduces a fully self-attention-based encoder that adaptively learns the topological and spatial  features. It lifts the dependence on RNN models.

\begin{figure}[ht]
  \centering
  \includegraphics[width=0.7\textwidth]{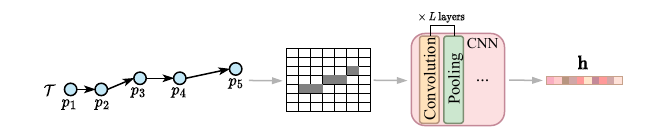}
  \caption{CNN-based learned measures}\label{fig:cnn}
\end{figure}

\textbf{Convolutional neural network (CNN)-based measures.}
CNN models are widely used in image representation learning. 
They stack convolution kernel and pooling layers to capture image features. 
To convert a trajectory to a fixed-size image, a blank image corresponding to the data space enclosing the trajectory is created. Then, the points on the trajectory are mapped to pixels of the image, where a pixel value indicates the number of points mapped to the pixel (cf.~Figure~\ref{fig:cnn}). Such a conversion resembles the grid-cell based  representation described earlier and shares similar benefits. 

\emph{TrjSR}~\cite{trjsr} is the only CNN-based model. It is trained by  reconstructing a super-resolution trajectory image from a low-resolution one. 
It encodes a trajectory in $\Omega(m{k^2}{n_k}c)$, where $k$ is the side length of the convolution kernels, $n_k$ is the number of kernels, $c$ is the channel size, and $m \gg d$ is the number of pixels. 
It takes $\Omega(k^2{n_k}c+mc)$ space, where the model parameters take $\Omega(k^2{n_k}c)$ space, and the intermediate results take $\Omega(mc)$ space.

An issue with the CNN-based measures is that they  lose the sequence information of the trajectory points. They cannot distinguish two trajectories traveling towards opposite directions.

\textbf{Graph neural network (GNN)-based measures.}
GNNs are designed for graph representation learning. 
In a typical GNN layer, every graph node receives and aggregates information (typically node embeddings) from its neighbors (aggregation). Then, the aggregated information is combined with the embedding of the node to form its updated embedding (combination).

\begin{figure}[ht]
  \centering
  \includegraphics[width=0.7\textwidth]{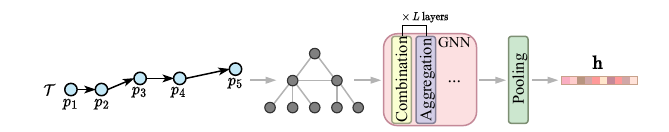}
  \caption{GNN-based learned measures}\label{fig:gnn}
\end{figure}

TrajGAT~\cite{trajgat} is the only GNN-based  measure  (cf.~Figure~\ref{fig:gnn}).
It builds a multi-level quadtree~\cite{quadtree} that partitions the space into cells with different sizes to help create graphs with multi-granularity views of trajectories.
To construct a graph from a trajectory, it queries each trajectory point in the quadtree  and adds the tree nodes on the traversed path into the graph as graph nodes. The edges between the added tree nodes are kept as graph edges. 
To encode a trajectory graph, TrajGAT takes $\Omega(n{n_e}d)$ time, where $n_e$ is the number of  neighbors per node in the graph. 
TrajGAT takes $\Omega(d^2+nd+n{n_e})$ space, where the model parameters take $\Omega(d^2)$ space, and the intermediate results for computing the node embeddings and the attention coefficients between nodes take $\Omega(nd+n{n_e})$ space.

\textbf{Discussion.} 
We have focused on  trajectories in Euclidean space. A closely related topic, trajectory similarity measures for road network space, has also attracted a strong  interest~\cite{trembr,gts,gts+,sarn,grlstm,lightpath,tsjoin,TP,dison,torch}. Both non-learned and learned measures have been proposed. The learned measures generally adopt GNNs to exploit the network connectivity patterns and learn embeddings for (intersections or road segments of) the road networks. A trajectory is represented by aggregating the embeddings of the intersections and/or road segments passed by it. Due to the rich works on this topic, we leave an empirical analysis for future work.

\subsection{Trajectory Similarity Queries}\label{subsec:related_knn}
Trajectory similarity queries typically refer to trajectory $k$ nearest neighbor ($k$NN) queries (Section~\ref{sec:preliminaries}).
All existing trajectory $k$NN query algorithms are designed for the non-learned measures.
Basically, they query trajectories from tree-based indies (e.g., R-trees~\cite{rtree}), with spatial distance-based pruning to speed up the search.

Most of the non-learned measures have their dedicated indices and query algorithms. 
LCSS comes with a 1-NN query algorithm~\cite{lcss}. It computes a distance lower bound between the query trajectory and the data trajectories in an index node. If this lower bound is greater than the distance between the query trajectory and the currently found nearest trajectory, the node (and all associated trajectories) can be safely pruned. 
EDR introduces several string encoding-based pruning techniques~\cite{edr}, including Q-gram (i.e., sub-string abstracts), string embedding, and triangle inequality based pruning~\cite{edr_pruning}.
ERP computes a distance lower bound based on the summation of trajectory points to prune dissimilar trajectories~\cite{erp}, while EDwP computes the distance between trajectories and bounding boxes of trajectories for pruning~\cite{edwp}.

There is also a generic R-tree-based structure for trajectory $k$NN queries~\cite{dtw_ts_index} that supports LCSS, DTW, and SPD. Its query algorithm creates a minimum bounding envelope (MBE) for the query trajectory, which bounds the search space by adding spatial and temporal offsets to the query trajectory. Trajectories outside the MBE are pruned.
\emph{DFT}~\cite{DFT} is a generic, distributed segment-based R-tree-like index that supports Hausdorff and Fr\'echet. 
The $k$NN algorithm of DFT leverages a small set of sampled data trajectories (more than $k$) to obtain a distance threshold $\epsilon$ that bounds the largest distance between the query trajectory $\mathcal{T}_q$ and its $k$-th NN. Then, DFT leverages $\epsilon$ to  filter out dissimilar data segments of $\mathcal{T}_q$.
\emph{DITA}~\cite{dita} is a third R-tree-based index, and it supports DTW and ERP. 
It indexes points sampled at equal intervals on each data trajectory, for space efficiency. 
Its $k$NN algorithm also starts by computing a $k$-th NN distance threshold $\epsilon$. 
Unlike DFT, DITA can shrink $\epsilon$ during its search process, by subtracting the distance between the currently matched points of a data trajectory and the query trajectory. 
This is because DTW and ERP compute trajectory distance by summing up the distances between the matched points. In contrast, Fr\'echet and Hausdorff compute the global maximum distance of the matched points, such that $\epsilon$ cannot be updated progressively.
We use DFT and DITA for indexing in the experiments due to their generalizability.

\subsection{Trajectory Clustering}\label{subsec:related_clustering}
Earlier studies on trajectory clustering~\cite{traclus,trajgroup,trajcluster_chenlu,trajcluster_online} apply classic clustering algorithms, e.g., $k$-medoids, with non-learned  measures (e.g., LCSS), 
For example, Lee et al.~\cite{traclus} presents a partition-and-group framework for trajectory clustering. Li et al.~\cite{trajgroup} focus on finding top-$k$ clusters considering the cluster cardinality. Besides, a series of studies~\cite{trajcluster_chenlu,trajcluster_online,trajgroup} consider online trajectory clustering.

Recent studies~\cite{trajclustering_yaodi,detect,trajcluster_olive,trip2vec,trajcluster_airspace,e2dtc} exploit deep learning  to improve clustering efficiency and effectiveness.
Yao et al.~\cite{trajclustering_yaodi} leverage an RNN auto-encoder. They first learn trajectory embeddings and then cluster the embeddings by the k-medoids algorithm. This clustering paradigm is followed by the later studies. 
\emph{DETECT}~\cite{detect} and \emph{Trip2Vec}~\cite{trip2vec} further consider POIs on trajectories to study the mobility pattern of trajectories.
\emph{E2DTC}~\cite{e2dtc} adopts t2vec~\cite{t2vec} as the encoder and  fine-tunes it with a multi-task loss function that considers both trajectory similarity and cluster distribution.
Besides, several studies~\cite{trajcluster_olive,trajcluster_ais} extend deep trajectory clustering to different data domains, e.g., aircraft or vessel trajectories.

\section{Experiments}\label{sec:exp}
We study the performance of both the non-learned and the learned measures empirically, for 
three representative tasks: trajectory similarity computation, clustering, and similarity querying. 

\subsection{Experimental Setup}\label{subsec:exp:setting}

\subsubsection{Datasets.} 
We use five real-world trajectory datasets, which are used in recent studies~\cite{neutraj,trjsr,t3s,trajgat,tmn,cl-tsim,trajcl}.
We pre-process the datasets by discarding short trajectories with less than 20 points, as done in previous studies~\cite{neutraj,tmn,trajcl,trjsr,t2vec,rsts}. 
Table~\ref{tab:dataset_statistics} summarizes dataset statistics after pre-processing.

\begin{table}[ht]
\centering
\caption{Dataset statistics}
\label{tab:dataset_statistics}
\setlength{\tabcolsep}{1mm}
\renewcommand{\arraystretch}{1.1}
\resizebox{0.75\columnwidth}{!}{
\begin{tabular}{l|c|c|c|c|c}
\hlineB{3}
 & \textbf{Porto} & \rvcolor{\textbf{Germany}} & \textbf{Geolife} & \rvcolor{\textbf{Chengdu}} & \textbf{Xi'an} \\ \hline \hline
\textbf{\#traj.} & 1,380,777 & \rvcolor{243,417} & 15,972 & \rvcolor{1,259,639} & 1,009,693   \\
\textbf{\#points per traj.} & 20 $\sim$ 3,836 &  \rvcolor{20 $\sim$ 200} & 20 $\sim$ 56,780 & \rvcolor{20 $\sim$ 1,891} & 20 $\sim$ 8,730   \\
\textbf{Avg. \#points per traj.} & 50 & \rvcolor{84} & 1,201 & \rvcolor{142} & 262    \\
\textbf{Traj. length (m)} & 8 $\sim$ 110,227 & \rvcolor{10,805 $\sim$ 1.4e8} & 1 $\sim$ 556,814 & \rvcolor{36 $\sim$ 51,789} & 23 $\sim$ 219,824   \\ 
\textbf{Avg. traj. length (m)} & 6,445 & \rvcolor{338,622} & 25,175 & \rvcolor{4,586} & 6,204   \\
\textbf{Spatial area (km$^2$)} & 16.0 $\times$ 20.1 & \rvcolor{1,023 $\times$ 1,187} & 235.9 $\times$ 232.5 & \rvcolor{9.7 $\times$ 9.7} & 9.9 $\times$ 9.6   \\
\textbf{Time span} & - & \rvcolor{-} & five years & \rvcolor{a week} &  a week  \\ \hlineB{3}
\end{tabular}
}
\end{table}

\textbf{Porto}~\cite{porto} contains taxi trajectories collected from Porto, Portugal, between July 2013 and June 2014. The trajectory points do not come with timestamps. The average number of points per trajectory is 50, which is the smallest among the five datasets.

\textbf{Germany}~\cite{osmplanet} contains user trajectories collected mainly within Germany from OpenStreetMap without timestamps. This dataset has the longest average trajectory length, i.e., 338 km, and covers the largest spatial area, i.e., over 10$^6$ km$^2$.

\textbf{Geolife}~\cite{geolife} contains user trajectories collected in Beijing, China, from April 2007 to August 2012. We have further discarded the trajectories recorded outside Beijing, which take up a very small portion (6\%) of the data. Geolife records trajectories of different types of user movements, e.g., walking, cycling, and driving. We use all trajectories together due to the limited dataset size.

\textbf{Chengdu}~\cite{didi} contains ride-hailing trajectories from the Second Ring road of Chengdu, China, which is a densely populated area, in the first week of November 2016. It covers the smallest area among the five datasets and has the shortest average trajectory length.

\textbf{Xi'an}~\cite{didi} contains ride-hailing trajectories from the Second Ring Road of Xi'an, China, in the first week of October 2018. This is the most current dataset and we use it by default. Its sampling rate is the lowest, i.e., roughly 30 meters per sampled point.

\subsubsection{Similarity measures.}
For each category of measures in Section~\ref{sec:relatedwork}, we study the state-of-the-art as well as measures that have unique computation strategies, such as NEUTRAJ and TMN, which are both RNN-based but use single- and dual-branch models, respectively. 
As there is no released code for SAX on trajectories, we follow the proposal~\cite{sax_traj1} and compute the duration for which two trajectories stay less than a given distance threshold based on their symbolic representations. We denote this measure as \emph{SAR}.
The similarity measures tested are listed below, where ``ST'' refers to measures that consider both spatial distances and time differences.

(1) Non-learned: \textcircled{\raisebox{-0.6pt}{1}} Linear scan:\: 
    {\small \textbf{CDDS} (ST) and \textbf{SAR} (ST)}
    \textcircled{\raisebox{-0.6pt}{2}}~DP:\: 
    {\small \textbf{DTW}, \textbf{ERP}, \textbf{Fr\'echet}, and \textbf{STEDR} (ST)}
    \textcircled{\raisebox{-0.6pt}{3}} Enumeration:\: 
    {\small \textbf{Hausdorff}}.
    
(2) Learned:
    \textcircled{\raisebox{-0.6pt}{1}} RNN:\: 
    {\small \textbf{NEUTRAJ} (single-branch), \textbf{TMN} (dual-branch), \textbf{RSTS} (ST)}
    \textcircled{\raisebox{-0.6pt}{2}} Self-attention:\: 
    {\small \textbf{T3S} (RNN $+$ self-attention), \textbf{TrajCL} (self-attention)}
    \textcircled{\raisebox{-0.6pt}{3}} CNN:\: 
    {\small \textbf{TrjSR}}
    \textcircled{\raisebox{-0.6pt}{4}}~GNN:\: 
    {\small \textbf{TrajGAT}}.

\subsubsection{Implementation details.}\label{subsubsec:exp:implementation}
All experiments are run on a virtual machine with an 8-core Intel Xeon 4214 CPU (2.2 GHz), 128GB RAM and an NVIDIA V100 GPU (5,120 FP32 cores and 16GB VRAM). 
We use these due to their ease of access for reproducibility reasons. 
We repeat each experiment five times and report the average results.

For the non-learned measures, we use \emph{traj-dist}~\cite{traj-dist}, a commonly used CPU-based sequential implementation of trajectory similarity computation in Python. 
It supports DTW, ERP, Fr\'echet, and Hausdorff, while we add support for STEDR, CDDS, and SAR.

We implement the GPU-based non-learned measures by following the meta-algorithms in Appendix~\ref{sec:implementation} on CUDA cores of GPUs in Python with the Numba 0.53.1 library supporting GPU programming. 
We note that parallel implementations of some of the non-learned measures exist (e.g.,~\cite{6831932}). We use our meta-GPU-based algorithms instead because these do not contain special optimizations for any specific measures, enabling us to capture better the speedups achievable by a simple GPU-based adaptation.

For the learned measures, we use their released source code that are written in PyTorch, which can be run on either CPU or GPU, 
except for T3S and RSTS for which no code is available. We implement T3S and RSTS with PyTorch~1.8.1 following their papers.

Following previous studies~\cite{neutraj,tmn,t3s}, we set the embedding dimensionality $d$ to 128. The side length of the grid cells is 100 meters for the grid-based learned measures. The image size $m$ of TrjSR is $162 \times 128$ following its paper.
The number of encoder layers stacked in each learned measure is 2, except for TrjSR, which has 10 CNN layers following its paper. We set the number of GPU cores, $n_c$, for computing the similarity between a pair of trajectories to 64. The batch size is 512 by default, to fit every measure in memory.

\begin{table}[]
\centering
\caption{Elapsed time of trajectory similarity computation (best results are in bold)}
\label{tab:comp_overall}
\setlength{\tabcolsep}{1mm}
\renewcommand{\arraystretch}{0.95}
\resizebox{0.55\columnwidth}{!}{
\begin{tabular}{l|ll|rr|rr}
\hlineB{3}
\multirow{2}{*}{\textbf{Dataset}} & \multicolumn{2}{c|}{\multirow{2}{*}{\textbf{Measure}}} & \multicolumn{2}{c|}{\textbf{Single}} & \multicolumn{2}{c}{\textbf{Batched}} \\ \cline{4-7} 
 & & & \multicolumn{1}{c}{\textbf{GPU}} & \multicolumn{1}{c|}{\textbf{CPU}} & \multicolumn{1}{c}{\textbf{GPU}} & \multicolumn{1}{c}{\textbf{CPU}} \\ \hline \hline
 
\multirow{10}{*}{\textbf{Porto}} & \multirow{4}{*}{{Non-learned}} & DTW & {146.23} & 36.04 & \textbf{3.96} & 10.07 \\
 &  & ERP & 190.76 & 78.46 & 4.09 & 12.88 \\
 &  & Fr\'echet & 146.90 & 32.59 & 4.00 & 7.44 \\
 &  & Hausdorff & \textbf{135.71} & \textbf{26.64} & 4.07 & \textbf{6.57} \\ \cdashline{2-7} 
 &  \multirow{6}{*}{{Learned}} & NEUTRAJ &  4907.14 & 3088.88 & 94.69 & 229.21  \\
 & & TMN &  389.63 & 535.48 & 54.67 & 254.23  \\
 & & T3S &  465.75 & 861.54 & 49.49 & 649.17  \\
 & & TrajCL &  561.94 & 666.82 & 59.49 & 276.30  \\ 
 & & TrjSR &  584.21 & OT & 301.74 & 4409.75  \\
 & & TrajGAT &  3319.11 & 2580.19 & 494.34 & 603.32  \\ \hline \hline
 
\multirow{10}{*}{\rvcolor{\textbf{Germany}}} & \multirow{4}{*}{\rvcolor{Non-learned}} & \rvcolor{DTW}  & \rvcolor{132.21} & \rvcolor{75.12} & \rvcolor{8.75} & \rvcolor{21.38} \\
 &  & \rvcolor{ERP}  & \rvcolor{163.34} & \rvcolor{191.38} & \rvcolor{\textbf{8.33}} & \rvcolor{35.03} \\
 &  & \rvcolor{Fr\'echet}  & \rvcolor{130.49} & \rvcolor{67.14} & \rvcolor{8.37} & \rvcolor{18.32} \\
 &  & \rvcolor{Hausdorff}  & \rvcolor{\textbf{113.22}} & \rvcolor{\textbf{39.21}} & \rvcolor{8.36} & \rvcolor{\textbf{14.42}} \\ \cdashline{2-7} 
 &  \multirow{6}{*}{\rvcolor{Learned}} & \rvcolor{NEUTRAJ}  & \rvcolor{OT} & \rvcolor{5893.80} & \rvcolor{164.62} & \rvcolor{367.51} \\
 &  & \rvcolor{TMN}  & \rvcolor{429.18} & \rvcolor{776.21} & \rvcolor{113.06} & \rvcolor{432.96} \\
 &  & \rvcolor{T3S}  & \rvcolor{428.82} & \rvcolor{979.93} & \rvcolor{74.93} & \rvcolor{392.11} \\
 &  & \rvcolor{TrajCL}  & \rvcolor{636.30} & \rvcolor{945.13} & \rvcolor{93.91} & \rvcolor{436.90} \\ 
 &  & \rvcolor{TrjSR}  & \rvcolor{914.14} & \rvcolor{OT} & \rvcolor{165.69} & \rvcolor{1909.11} \\
 &  & \rvcolor{TrajGAT}  & \rvcolor{2440.17} & \rvcolor{2194.09} & \rvcolor{588.12} & \rvcolor{1775.44} \\  \hline \hline
 
 \multirow{14}{*}{\textbf{Geolife}} & \multirow{4}{*}{{Non-learned}} & DTW & 165.68 & 130.19 & \textbf{5.68} & 24.70 \\
 &  & ERP &  212.14 & 283.55 & 5.94 & 48.91  \\
 &  & Fr\'echet &  165.91 & 91.40 & 5.82 & 23.02  \\
 &  & Hausdorff &  \textbf{140.97} & \textbf{47.29} & 5.76 & \textbf{16.95}  \\ \cdashline{2-7} 
 &  \multirow{6}{*}{{Learned}} & NEUTRAJ & 4182.19 & 2682.02 & 118.14 & 306.60  \\
 &  & TMN &  524.18 & 608.16 & 68.64 & 414.70  \\
 &  & T3S &  618.42 & 1242.94 & 93.35 & 833.46  \\
 &  & TrajCL &  613.15 & 711.60 & 88.28 & 425.05 \\ 
 &  & TrjSR & 588.39 & OT & 299.55 & 4417.78 \\
 &  & TrajGAT & 4375.38 & 3799.01 & 1335.27 & 1815.51 \\ \cline{2-7} 
 & \multirow{3}{*}{\begin{tabular}[c]{@{}l@{}}Non-learned\\ (ST)\end{tabular}} & {STEDR} & 165.46 & 221.35 & \textbf{6.50} & 43.94 \\
 &   & {CDDS} & \textbf{138.17} & \textbf{24.74} & 6.65 & \textbf{17.35} \\
 &   & \rvcolor{SAR} & \rvcolor{158.97} & \rvcolor{30.45} & \rvcolor{6.85} & \rvcolor{17.53}  \\ \cdashline{2-7} 
 &  Learned (ST) & {RSTS} & 2970.58 & 3745.92 & 805.44 & 939.33 \\ \hline \hline
 
\multirow{14}{*}{\rvcolor{\textbf{Chengdu}}} & \multirow{4}{*}{\rvcolor{Non-learned}} & \rvcolor{DTW}  & \rvcolor{139.91} & \rvcolor{112.59} & \rvcolor{9.87} & \rvcolor{29.34}  \\
 &  & \rvcolor{ERP}  & \rvcolor{168.70} & \rvcolor{298.32} & \rvcolor{9.55} & \rvcolor{53.43}   \\
 &  & \rvcolor{Fr\'echet}  & \rvcolor{138.66} & \rvcolor{102.87} & \rvcolor{\textbf{9.50}} & \rvcolor{25.56}  \\
 &  & \rvcolor{Hausdorff}  & \rvcolor{\textbf{116.74}} & \rvcolor{\textbf{50.56}} & \rvcolor{9.55} & \rvcolor{\textbf{18.78}} \\ \cdashline{2-7} 
 &  \multirow{6}{*}{ \rvcolor{Learned} } & \rvcolor{NEUTRAJ}  & \rvcolor{5235.88} & \rvcolor{3306.81} & \rvcolor{108.83} & \rvcolor{222.66}  \\
 &  & \rvcolor{TMN}  & \rvcolor{251.70} & \rvcolor{409.24} & \rvcolor{78.08} & \rvcolor{220.05} \\
 &  & \rvcolor{T3S}  & \rvcolor{477.05} & \rvcolor{1188.18} & \rvcolor{89.10} & \rvcolor{428.20} \\
 &  & \rvcolor{TrajCL}  & \rvcolor{617.49} & \rvcolor{749.70} & \rvcolor{81.66} & \rvcolor{229.45}  \\ 
 &  & \rvcolor{TrjSR}  & \rvcolor{947.35} & \rvcolor{OT} & \rvcolor{197.99} & \rvcolor{1950.42}  \\
 &  & \rvcolor{TrajGAT}  & \rvcolor{5713.50} & \rvcolor{5387.81} & \rvcolor{2700.47} & \rvcolor{5682.02} \\ \cline{2-7} 
 &  \multirow{3}{*}{\begin{tabular}[c]{@{}l@{}} \rvcolor{Non-learned} \\ \rvcolor{(ST)} \end{tabular}} & \rvcolor{STEDR}  & \rvcolor{137.57} & \rvcolor{233.34} & \rvcolor{12.11} & \rvcolor{47.82} \\
 &  & \rvcolor{CDDS}  & \rvcolor{\textbf{114.57}} & \rvcolor{30.18} & \rvcolor{\textbf{11.85}} & \rvcolor{20.20} \\ 
 &  & \rvcolor{SAR} & \rvcolor{160.68} & \rvcolor{\textbf{29.79}} & \rvcolor{12.69} & \rvcolor{\textbf{17.53}} \\ \cdashline{2-7} 
 &  \rvcolor{Learned (ST)} & \rvcolor{RSTS}  & \rvcolor{1964.53} & \rvcolor{4409.20} & \rvcolor{1087.39} & \rvcolor{1212.71} \\ \hline \hline

\multirow{14}{*}{\textbf{Xi'an}} & \multirow{4}{*}{{Non-learned}} & DTW & 168.24 & 133.42 & 6.21 & 28.11 \\
 &  & ERP & 215.28 & 362.92 & 6.34 & 54.04 \\
 &  & Fr\'echet & 169.60 & 117.20 & 6.49 & 25.66 \\
 &  & Hausdorff & \textbf{145.41} & \textbf{70.63} & \textbf{5.83} & \textbf{17.75} \\ \cdashline{2-7} 
 &  \multirow{6}{*}{{Learned}} & NEUTRAJ & 4226.71 & 2609.39 & 102.85 & 215.03 \\
 &  & TMN & 394.89 & 463.79 & 62.08 & 230.18 \\
 &  & T3S & 639.53 & 1958.81 & 83.69 & 778.61 \\
 &  & TrajCL & 622.56 & 733.51 & 71.37 & 298.11 \\ 
 &  & TrjSR & 625.54 & OT & 318.97 & 4629.79 \\
 &  & TrajGAT & 4439.46 & 4114.52 & 1112.02 & 1560.24 \\ \cline{2-7} 
 &  \multirow{3}{*}{\begin{tabular}[c]{@{}l@{}}Non-learned\\ (ST)\end{tabular}} & {STEDR} & 171.03 & 280.66 & 7.25 & 48.09 \\
 &   & {CDDS} & \textbf{142.51} & \textbf{33.77} & \textbf{7.06} & 22.66 \\ 
 &  & \rvcolor{SAR} & \rvcolor{162.54} & \rvcolor{34.27} & \rvcolor{7.22} & \rvcolor{\textbf{19.40}} \\ \cdashline{2-7} 
 &   Learned (ST) & {RSTS} & 3736.67 & 6030.93 & 1257.75 & 1473.71 \\ 
 \hlineB{3}
\end{tabular}
}
\end{table}

\subsection{Trajectory Similarity Computation}\label{subsec:exp:trajsimi}

We first report the results on trajectory similarity computation.

\subsubsection{Setup.}\label{subsubsec:exp:trajsimi_setup}
For each dataset, we randomly sample 100,000 pairs of trajectories and compute their similarity using each measure. Each trajectory is limited to at most 200 points, following previous studies~\cite{neutraj,trajgat,tmn,trajcl,rsts}, as the learned measures may fail on longer trajectories due to out-of-memory errors (cf.~Section~\ref{subsubsec:exp:trajsimi_numpoints}). 
\textit{Each trajectory is used only once, and no embeddings can be reused in Section~\ref{subsec:exp:trajsimi} by default.} This experiment setting aims to evaluate the efficiency of computing the similarity of a pair of trajectories arriving online.
We vary the data size of each run (i.e., \textbf{single} or \textbf{batched}) and the computation unit (i.e., \textbf{CPU} or \textbf{GPU}), to provide comprehensive results that cover different application settings.
Here, ``single'' refers to computing for a single pair of trajectories (with a single core), while ``batched'' refers computing for multiple pairs (i.e., 512) of trajectories using multiple cores of a computation unit.

Note that, when we compute trajectory similarity in batches,  different ``batched'' computation paradigms are applied on GPU and CPU, respectively. On GPU, we use $n_c$ cores for parallel processing of each trajectory pair following our meta-GPU-based algorithms. 
On CPU, we simply use a computation core for each trajectory pair, i.e., no parallel processing is done on individual trajectory level. This is because there are much more cores on GPU than on CPU~\cite{ward2014real}. In batch processing mode, all CPU cores can be fully utilized by the trajectory pairs in a batch already, while the cores on GPU can be shared at the individual trajectory level.   

We run on both CPU and GPU for two reasons: (1)~to show the speedups achievable using GPUs for the \emph{same} trajectory similarity measure and guide the choice between CPU (less expensive but slower) and GPU (faster but more expensive) for different applications; and (2) to compare the speedups achievable using GPUs for \emph{different} trajectory similarity measures, to reveal the measures that can better exploit the power of GPU parallelization.

We report the \textbf{elapsed time} (in seconds) and the \textbf{space cost} (in GB). 
All input data is aligned to the same form, i.e., raw 2-dimensional trajectory points, and the outputs are similarity scores. 
The cutoff running time is 7,200 seconds. We use  ``\textbf{OT}'' to denote overtime errors and ``\textbf{OOM}'' to denote out-of-memory errors.

\subsubsection{Results under Online Computation Settings.}\label{subsubsec:exp:trajsimi_overall} 
Table~\ref{tab:comp_overall} shows the elapsed times when all computation is done online, including the trajectory embeddings.
\textit{Overall, the non-learned measures take less time to compute than the learned ones when they are run online on the same computation units.} 
Even for batched computation on GPU, where the learned ones are supposed to be at their best, the non-learned measures are at least an order of magnitude faster. 

A few detailed observations can be made from the table:

(1) \textbf{Both non-learned and learned measures take the least time for batched computation on GPU.} This setting best exploits  the parallelization power of GPU. Such a setting thus should be used for one-off similarity computation of a large number of trajectories (e.g., for offline trajectory mining tasks such as contact tracing).  

(2) \textbf{When computing the similarity between trajectories on per pair basis (``single''), e.g., for an ad hoc similarity computation, 
the non-learned measures prefer CPU while the learned ones prefer GPU.} 
The non-learned measures are lightweight, such that the savings achieved by GPU for a single trajectory pair is not worth the data transfer costs between CPU and GPU. 
The learned measures, on the other hand, take less time on GPU, where matrix multiplications in the trajectory encoders are better suited.

(3) \textbf{Among the non-learned measures, the enumeration-based one, Hausdorff, is the fastest in general} (excluding the spatio-temporal measures),  due to its simple computation rules. When computed on GPU, Hausdorff further benefits from its independent and fully parallelized calculation of the pairwise point similarity scores. Among the  DP-based measures, ERP generally takes the most time, as it has the most complex computation rules.  We note that all non-learned measures are very fast and have similar elapsed times under the GPU-batched mode, while DTW reported marginally faster elapsed time on Porto and Geolife. 

Among the spatio-temporal measures, the linear scan-based CDDS is generally the fastest on GPU for its simple calculation. 
SAR performs similar to CDDS which is also a linear-time measure. On CPU, SAR can run faster than CDDS, because it can early terminate the computation when two trajectories do not align in time. This early termination does not help on GPU as the trajectories are distributed to each streaming core for processing anyway.

(4) \textbf{Among the learned measures, the attention-based ones T3S and TrajCL are more efficient in time} -- T3S is the slower among the two as it also uses an RNN. This is because the attention computation can be fully paralellized (cf.~Section~\ref{sec:relatedwork}). An RNN-based measure, TMN, is the fastest among the learned measures, due to its simple model (a vanilla LSTM). NEUTRAJ and RSTS are also RNN-based, while they suffer in efficiency especially on the ``single'' mode. NEUTRAJ has an expensive spatial module to compute the attention coefficients between the current and the seen training trajectories (cf. Section~\ref{sec:relatedwork}), while 
RSTS has an expensive input pre-processing step (cf. Table~\ref{tab:comp_detail_time_pairs}). 
The CNN-based measure TrjSR and the GNN-based measure TrajGAT are also slow. TrjSR suffers in the large amount of computation for its CNN, while TrajGAT spends much time on converting a trajectory into a graph.

\begin{table}[htp]
\centering
\caption{Detailed time and space costs of batched trajectory similarity computation on Xi'an}
\label{tab:comp_detail_time_pairs}
\setlength{\tabcolsep}{0.5mm}
\renewcommand{\arraystretch}{1.1}
\resizebox{0.7\columnwidth}{!}{
\begin{tabular}{l|rrrr|rrrr|r}
\hlineB{3} 
\multirow{2}{*}{\textbf{Measure}} & \multicolumn{4}{c|}{\textbf{Time on GPU}} & \multicolumn{4}{c|}{\textbf{Time on CPU}} & \multicolumn{1}{c}{\multirow{2}{*}{\begin{tabular}[c]{@{}r@{}}\textbf{Space} \\ (GB)\end{tabular}}}\\ \cline{2-9} 
 & \multicolumn{1}{c}{\textbf{Pre.}} & \multicolumn{1}{c}{\textbf{Emb.}} & \multicolumn{1}{c}{\textbf{Cmp.}} & \multicolumn{1}{c|}{\textbf{Total}} & \multicolumn{1}{c}{\textbf{Pre.}} & \multicolumn{1}{c}{\textbf{Emb.}} & \multicolumn{1}{c}{\textbf{Cmp.}} & \multicolumn{1}{c|}{\textbf{Total}} & \\ \hline \hline
DTW & 5.64 & - & 0.56 & 6.21 & 0.65 & - & 27.45 & 28.11 & \textbf{0.02} \\
ERP & 5.72 & - & 0.62 & 6.34 & 0.52 & - & 53.51 & 54.04 & \textbf{0.02} \\
Fr\'echet & 5.91 & - & 0.58 & 6.49 & 0.50 & - & 25.16 & 25.66 & \textbf{0.02}  \\
Hausdorff & 5.28 & - & 0.55 & \textbf{5.83} & 0.44 & - & 17.30 & \textbf{17.75} & \textbf{0.02} \\ \cdashline{1-10} 
NEUTRAJ & 68.15 & 34.69 & 1e-4 & 102.85 & 67.83 & 147.20 & 2e-3 & 215.03 & 0.13 \\ 
TMN & 56.10 & 5.97 & 1e-4 & 62.08 & 56.18 & 174.00 & 2e-3 & 230.18 & 6.37 \\
T3S & 59.73 & 23.95 & 1e-4 & 83.69 & 58.27 & 720.34 & 2e-3 & 778.61 & 3.02 \\
TrajCL & 60.38 & 10.99 & 1e-4 & 71.37 & 57.79 & 240.30 & 2e-3 & 298.11 & 5.81  \\
TrjSR & 86.56 & 232.40 & 1e-4 & 318.97 & 90.08 & 4539.71 & 2e-3 & 4629.79 & 9.12 \\
TrajGAT & 1035.67 & 76.35 & 1e-4 & 1112.02 & 1028.66 & 531.58 & 2e-3 & 1560.24 & 13.89 \\ \hline
{STEDR} (ST) & 6.72 & - & 0.53 & 7.25 & 0.83 & - & 47.26 & 48.09 & \textbf{0.02} \\
{CDDS} (ST) & 6.49 & - & 0.57 & \textbf{7.06} & 0.60 & - & 22.06 & 22.66 & \textbf{0.02} \\ 
\rvcolor{{SAR} (ST)} &  \rvcolor{6.60} & \rvcolor{-} & \rvcolor{0.62} & \rvcolor{7.22} & \rvcolor{0.68} & \rvcolor{-} & \rvcolor{18.72} & \rvcolor{\textbf{19.40}} & \rvcolor{0.03} \\ \cdashline{1-10} 
{RSTS} (ST) & 1239.22 & 18.53 & 1e-4 & 1257.75 & 1245.93 & 227.78 & 2e-3 & 1473.71 & 0.76 \\ \hlineB{3}
\end{tabular}
}
\vspace{1mm}
\end{table}

\textbf{Computation time decomposition.} 
We further ``zoom in'' on the time costs. 
The elapsed time of a similarity measure reported in Table~\ref{tab:comp_overall}  (\textbf{Total} in Table~\ref{tab:comp_detail_time_pairs}) mainly consists of three parts: 
(1)~input pre-processing time (\textbf{Pre.}) to convert a raw trajectory into the format required by a measure (e.g., a graph for TrajGAT), (2)~embedding time (\textbf{Emb.}) to compute trajectory embeddings (inapplicable to the non-learned measures), and (3)~similarity computation time (\textbf{Cmp.}), i.e., the time to compute point matches and distances for the non-learned measures, or to compute the embedding distances for the learned measures.
There is also time for transferring the results and other minor inter-step processing, which is very small and hence omitted from the table.  
We only show the results on Xi'an in Table~\ref{tab:comp_detail_time_pairs}, as similar patterns are observed on the other datasets.

Overall, although the non-learned measures take more time than the learned measures on similarity computation, they have better efficiency on input pre-processing and do not require trajectory embeddings, which explains for their smaller total elapsed times.
The non-learned measures take more time for input pre-processing on GPU than on CPU. This is because when preparing data on GPU, we need to pad the trajectories in a batch to the same length and group them into a matrix, which is required by CUDA.

Table~\ref{tab:comp_detail_time_pairs} also suggests that the learned measures can be faster (in Cmp.), when the embeddings can be pre-computed and reused.

\begin{table}[h]
\small
\centering
\caption{Maximum batch sizes for different measures on Xi'an}
\label{tab:batchsize}
\setlength{\tabcolsep}{1.5pt}
\renewcommand{\arraystretch}{1.1}
\resizebox{0.55\columnwidth}{!}{%
\begin{tabular}{rl@{\hspace{5pt}}:l@{\hspace{5pt}}rl@{\hspace{5pt}}:l@{\hspace{5pt}}ll}
\hlineB{3}
DTW: & 100,000 &  & NEUTRAJ: & 100,000 &  & \multicolumn{1}{r}{STEDR:} & 100,000 \\
ERP: & 100,000 &  & TMN: & 2,048 &  & \multicolumn{1}{r}{CDDS:} & 100,000 \\
Fr\'echet: & 100,000 &  & T3S: & 2,048 &  & \multicolumn{1}{r}{\rvcolor{SAR:}} & \rvcolor{100,000} \\
Hausdorff: & 100,000 &  & TrajCL: & 2,048 &  & \multicolumn{1}{r}{RSTS:} & 16,384  \\
\multicolumn{1}{l}{\textbf{}} &  &  & TrjSR: & 512 &  &  &  \\
\multicolumn{1}{l}{\textbf{}} &  &  & TrajGAT: & 512 &  &  &  \\ \hlineB{3}
\end{tabular}%
}
\end{table}
\textbf{Memory costs.}  Table~\ref{tab:comp_detail_time_pairs} further shows  that the non-learned measures take less memory, as they do not need large weight matrices.

We next show the maximum number of trajectory pairs that each measure can process in parallel given the same GPU memory (16 GB VRAM). We double the batch size from 256 until reaching 100,000 which is the number of trajectories in the dataset. 
Table~\ref{tab:batchsize} shows the results on Xi'an. 
Overall, the non-learned measures allow a larger batch size than the learned ones except NEUTRAJ which also has a good space efficiency (for its simple RNN). Even so, the non-learned measures take less memory than NEUTRAJ, i.e., 1 GB vs. 12 GB. 
TMN, T3S, TrajCL, and TrajGAT need to compute the correlation between every two points, while  TrjSR suffers from much intermediate computation, which cost more memory space.

\begin{figure}[htp]
    \subfloat[{Computation on GPU}~\label{fig:exp_comp_numpoints_gpu}]{
        \includegraphics[width=0.8\textwidth]{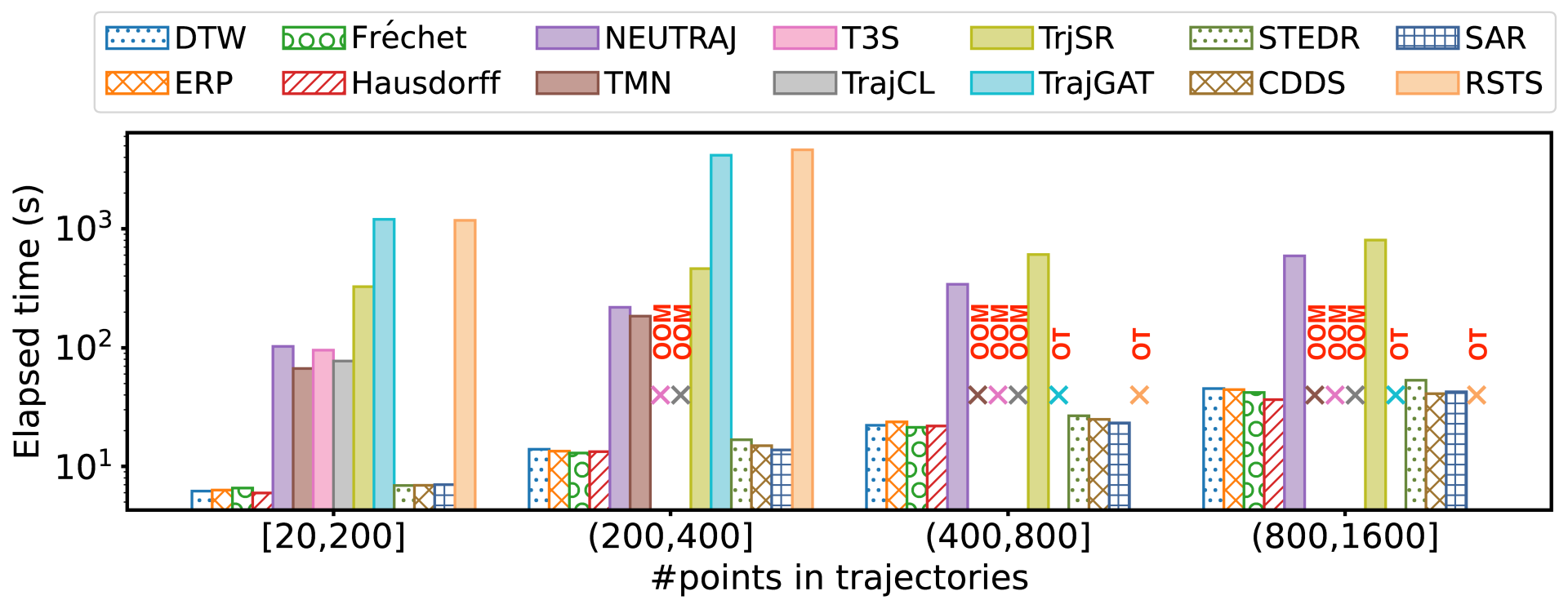}
    } \\
    \subfloat[{Computation on CPU}~\label{fig:exp_comp_numpoints_cpu}]{
        \includegraphics[width=0.8\textwidth]{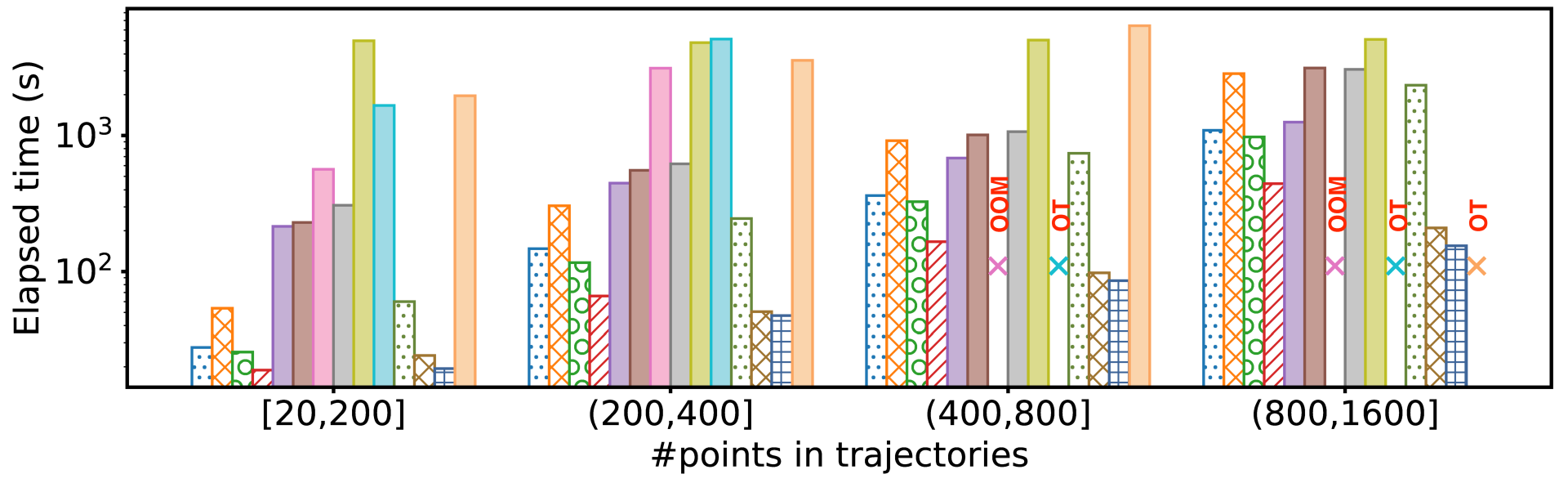}
    } \\ \vspace{1mm}
    \caption{Elapsed time vs. \#points in trajectories}\label{fig:exp_comp_numpoints}
\end{figure}

\subsubsection{Impact of the Number of Points in Trajectories $n$.}\label{subsubsec:exp:trajsimi_numpoints}
We vary the number of points on trajectories, $n$, from 20 to 1,600. Most existing studies on learned measures only used trajectories with up to 200 points. Ours  is the first set of results with over 1,000 points.

We focus on the batched setting on GPU and CPU as this is a more realistic setting in practice, and we omit the ``single''-mode results as similar result patterns are observed as before. We again present the results on the largest dataset, Xi'an, for brevity. 

Figure~\ref{fig:exp_comp_numpoints} shows the results, where each configuration, e.g., ``[20, 200]'' means to compute trajectory similarity for 100,000 pairs of randomly sampled trajectories each with $n = 20$ to $200$. 
All measures except TrjSR have increasing elapsed times as $n$ increases. The non-learned measures (denoted by empty bars with hatches) show clear advantage on GPU, while both types of measures perform closer on CPU when $n$ becomes larger. 
TrjSR uses fixed-size images to represent trajectories, which is independent from $n$. 
TMN, T3S, and TrajCL trigger out of memory errors when $n$ becomes large, especially on GPU. This is because TMN computes the attention coefficients between all pairs of points on two trajectories, while T3S and TrajCL compute the coefficients between every two points on each trajectory. These coefficient computations lead to a quadratic memory space overhead with respect to $n$.

\subsubsection{Impact of the Trajectory Embedding Dimensionality $d$}\label{subsubsec:exp:trajsimi_embdim}
\begin{figure}[htp]
        \includegraphics[width=0.8\textwidth]{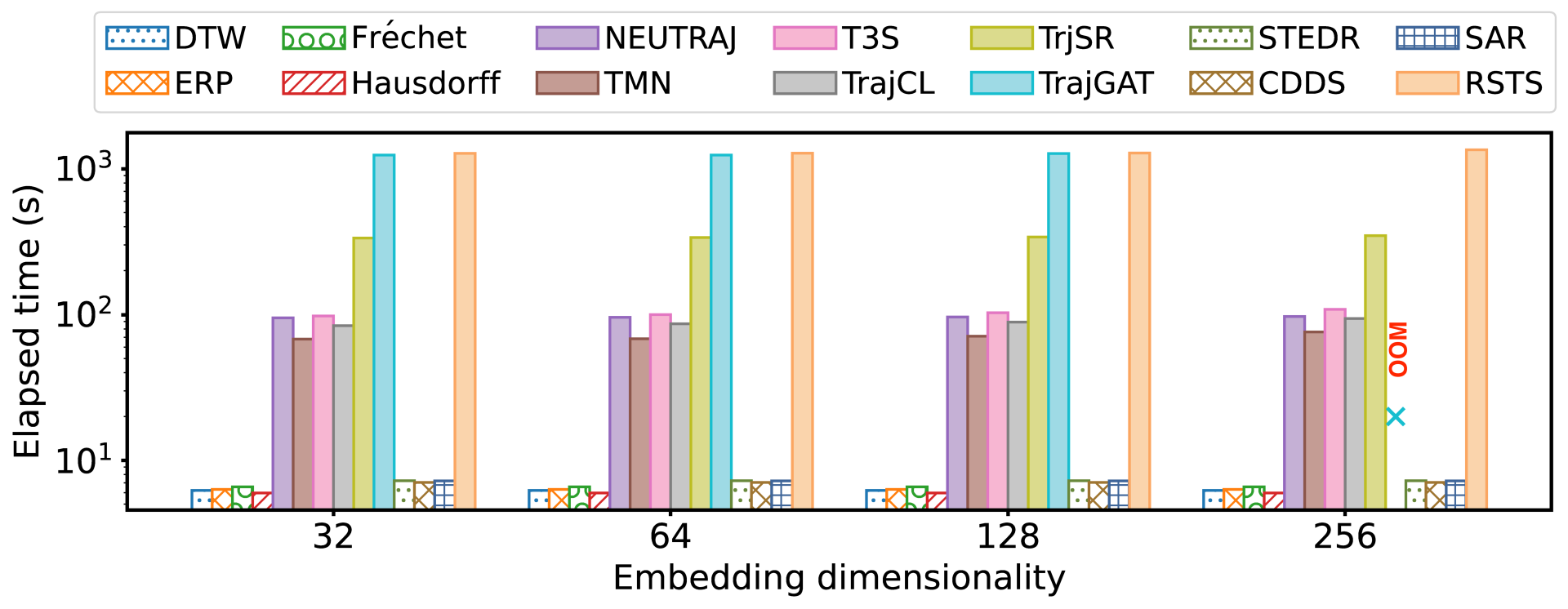}
    \caption{Elapsed time vs. embedding dimensionality (GPU)}\label{fig:exp_comp_embdim}
\end{figure}

We vary $d$ from 32 to 256 following the literature and report the results in Figure~\ref{fig:exp_comp_embdim}. 
The non-learned measures are not impacted by $d$.
The elapsed times of the learned measures present only a slightly increasing trend, as their matrix operations have been well parallelized by the PyTorch package. The learned measures are still slower even when $d = 32$, which is the smallest $d$ value used in the literature~\cite{trajgat}. 
When $d$ increases to 256, TrajGAT has an out of memory error, because it takes more space to compute the graph-based trajectory embeddings.
We show only the results on GPU for conciseness as similar patterns are observed on CPU.

\subsubsection{Impact of the Number of Trajectory Pairs}

We further vary the number of trajectory pairs from 1,000 to 1,000,000. Again, the computation time grows with the number of trajectory pairs, and the non-learned measures outperform the learned ones. These results reinforce the advantage of the non-learned measures in time efficiency. We omit the result figures for brevity.


\subsubsection{Impact of Computation Hardware.}\label{subsubsec:exp:trajsimi_hardware}
We also study the impact of different computation hardware on the efficiency of the similarity measures. We compare the default devices with Intel Xeon 6326 CPU (8 cores at 2.9 GHz) and NVIDIA A100 GPU (8,192 FP32 cores and 80 GB VRAM), which are more advanced processors. 
We repeat the trajectory similarity computation experiments on Xi'an. Table~\ref{tab:device} shows the results, which confirm that the similarity measures benefit from the more advanced computation hardware, as expected.

\begin{table}[htp]
\centering
\caption{Comparison of the time costs of trajectory similarity computation between different GPUs and CPUs}
\label{tab:device}
\renewcommand{\arraystretch}{1.1}
\resizebox{0.8\columnwidth}{!}{%
\begin{tabular}{l|rrrr|rrrr}
\hlineB{3}
\multirow{3}{*}{\textbf{Measure}} & \multicolumn{4}{c|}{\textbf{GPU (NVIDIA)}} & \multicolumn{4}{c}{\textbf{CPU (Intel Xeon)}} \\ \cline{2-9} 
 & \multicolumn{2}{c|}{\textbf{Single}} & \multicolumn{2}{c|}{\textbf{Batched}} & \multicolumn{2}{c|}{\textbf{Single}} & \multicolumn{2}{c}{\textbf{Batched}} \\ \cline{2-9} 
 & \multicolumn{1}{c|}{\textbf{V100}} & \multicolumn{1}{c|}{\textbf{A100} \faThumbsOUp} & \multicolumn{1}{c|}{\textbf{V100}} & \multicolumn{1}{c|}{\textbf{A100} \faThumbsOUp} & \multicolumn{1}{c|}{\textbf{4214}} & \multicolumn{1}{c|}{\textbf{6326} \faThumbsOUp} & \multicolumn{1}{c|}{\textbf{4214} } & \multicolumn{1}{c}{\textbf{6326} \faThumbsOUp} \\ \hline \hline
DTW & 168.24 & 122.50 & 6.21 & 3.45 & 133.42 & 92.52 & 28.11 & 25.74 \\
ERP & 215.28 & 153.66 & 6.34 & 3.53 & 362.92 & 260.84 & 54.04 & 46.17 \\
Frechet & 169.60 & 122.38 & 6.49 & 3.61 & 117.20 & 80.64 & 25.66 & 20.91 \\
Hausdorff & 145.41 & 122.55 & 5.83 & 3.24 & 70.63 & 43.22 & 17.75 & 15.31 \\ \cdashline{1-9}
NEUTRAJ & 4226.71 & 3308.69 & 125.07 & 86.05 & 2609.39 & 2089.93 & 215.03 & 179.38 \\ 
TMN & 394.89 & 205.74 & 62.08 & 57.70 & 463.79 & 326.25 & 230.18 & 190.64 \\
T3S & 639.53 & 419.97 & 83.69 & 74.83 & 1958.81 & 1149.50 & 778.61 & 440.76 \\
TrajCL & 622.56 & 512.66 & 71.37 & 65.36 & 733.51 & 585.01 & 298.11 & 204.18 \\
TrjSR & 625.54 & 373.38 & 318.97 & 127.28 & OT & OT & 4629.79 & 2281.05 \\
TrajGAT & 4439.46 & 3055.88 & 1112.02 & 974.56 & 4114.52 & 2818.22 & 1560.24 & 1273.82 \\ 
\hline
STEDR & 171.03 & 115.34 & 7.25 & 4.04 & 280.66 & 200.96 & 48.09 & 38.45 \\
CDDS & 142.51 & 93.45 & 7.06 & 3.93 & 33.77 & 25.29 & 22.66 & 14.18 \\
SAR & 162.54 & 109.46 & 7.22 & 4.02 & 34.27 & 20.73 & 19.40 & 13.62 \\ \cdashline{1-9}
RSTS & 3736.67 & 2048.62 & 1257.75 & 1114.00 & 6030.93 & 4154.37 & 1473.71 & 1168.49 \\ 
\hlineB{3}

\end{tabular}%
}
\end{table}

\subsubsection{Impact of the Number of GPU Cores $n_c$} \label{subsubsec:exp:trajsimi_numgpucores}

\begin{figure}[ht]
    \subfloat[{\#GPU cores per trajectory pair}~\label{fig:exp_num_gpu_cores}]{
            \includegraphics[width=0.35\textwidth]{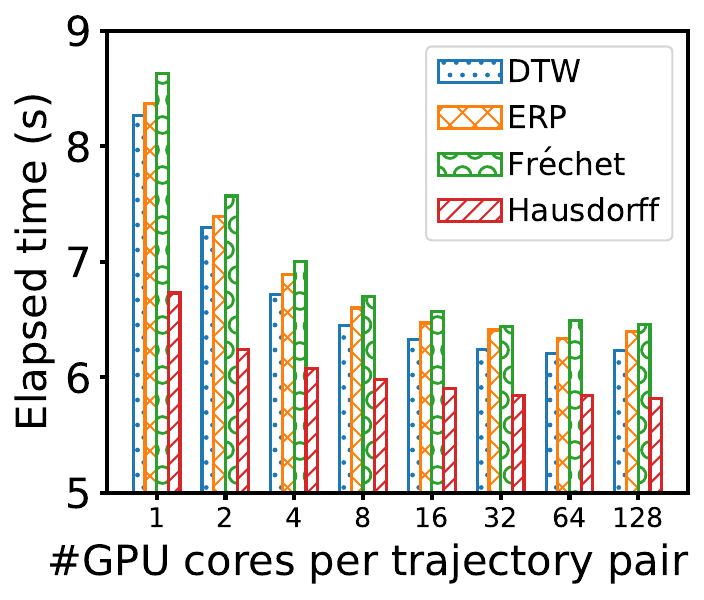}
    }
    \subfloat[{TrjSR image size}~\label{fig:exp_learned_params_trjsr}]{
        \includegraphics[width=0.20\textwidth]{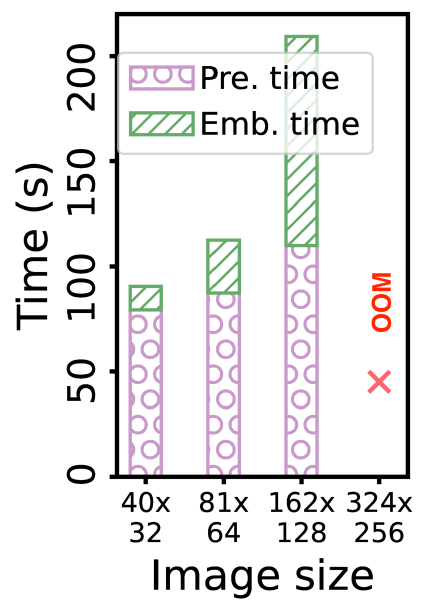}
    }
    \subfloat[{Quadtree node capacity of TrajGAT}~\label{fig:exp_learned_params_trajgat}]{
        \includegraphics[width=0.21\textwidth]{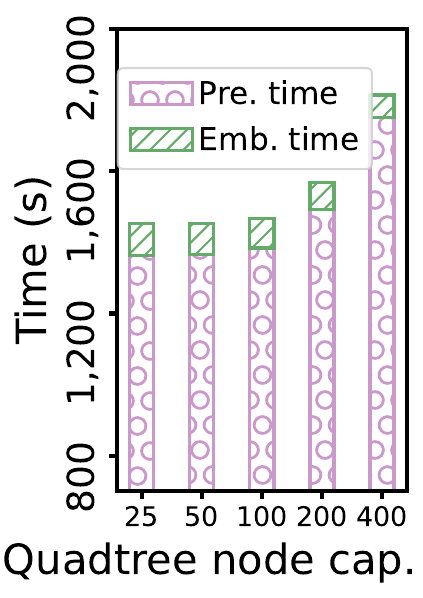}
    }
    \caption{Elapsed time vs. hardware and hyper-parameters}\label{fig:exp_params}
\end{figure}

We vary the number of GPU cores $n_c$ for computing the similarity between a pair of trajectories from 1 to 128. Figure~\ref{fig:exp_num_gpu_cores} reports the results on Xi'an for the non-learned measures (the learned measures are implemented based on PyTorch which does not offer control on the number of GPU cores for such computation). 
Overall, when $n_c$ increases, the elapsed times first decrease before stabilizing at around $n_c = 32$.  
This is because when $n_c$ grows, the number of trajectory points assigned to each GPU core decreases, which helps reduce the elapsed time. When $n_c$ grows further, the benefit of reducing the number of points per core shrinks, while the overhead on scheduling more cores increases, such that the elapsed time does not decrease further.

\subsubsection{Impact of the Hyper-parameters of Learned Measures.}\label{subsubsec:exp:trajsimi_learnedparams}
Hyper-parameters of the learned measures, such as the image size $m$ in TrjSR and the quadtree node capacity $q_{nc}$ in TrajGAT, also impact the time efficiency.
Particularly, the quadtree node capacity determines the quadtree structure in TrajGAT, which is inversely proportional to the number of neighbors of a graph node in TrajGAT, $n_e$ -- the time complexity of TrajGAT is linear in $n_e$ (cf. Section~\ref{subsec:related_learned}).

We vary $m$ and $q_{nc}$, and we report the detailed time costs for batched similarity computation on GPU in Figures~\ref{fig:exp_learned_params_trjsr} and~\ref{fig:exp_learned_params_trajgat}, respectively.
When $m$ increases, TrjSR takes more time to run, since the input size grows. 
For TrajGAT, when $q_{nc}$ increases, the pre-processing time increases, because its hash computation to identify quadtree node candidates becomes more complex during graph construction. 
The embedding time decreases, since each node corresponds to a larger area, and each trajectory can be represented with fewer nodes, and $n_e$ also decreases. Overall, the total time increases with $q_{nc}$, which is dominated by the pre-processing time.

\subsubsection{Results with Embedding Reuse.}~\label{subsubsec:exp:trajsimicomp_sets} 
Existing studies~\cite{trajcl,t3s,neutraj} assume two sets of trajectories and report the time to compute the similarity between each pair of trajectories, one from each dataset. \emph{They compute the embedding of each trajectory once and reuse it for all similarity computations}, which saves the overall elapsed time substantially.
We repeat such a set of experiments to show that, when the embeddings can be pre-computed, the learned measures indeed can delivery their promised higher computation efficiency.

On each dataset, we randomly sample and form two sets of 1,000 and 100 trajectories, respectively, such that we have 100,000 trajectory pairs to compute as before. 
We report the results of batched computation on Xi'an in Table~\ref{tab:comp_detail_time_QD}. 
Now the learned measures TrajCL, T3S, and NEUTRAJ outperform the non-learned ones, benefiting from the one-off input pre-processing and embedding times. TrjSR and TrajGAT also benefit substantially. However, they are still slower than the non-learned ones because even the one-off computations are too expensive. These results are consistent with the literature~\cite{trajcl}.
TMN does not apply in this experiment, as it is a pairwise model and cannot pre-compute for individual trajectories.

\begin{table}[htp]
\centering
\caption{Detailed time and space costs of batched  trajectory similarity computation between two trajectory sets on Xi'an}
\label{tab:comp_detail_time_QD}
\setlength{\tabcolsep}{0.5mm}
\renewcommand{\arraystretch}{1.1}
\resizebox{0.7\columnwidth}{!}{
\begin{tabular}{l|rrrr|rrrr|r}
\hlineB{3} 
\multirow{2}{*}{\textbf{Measure}} & \multicolumn{4}{c|}{\textbf{Time on GPU}} & \multicolumn{4}{c|}{\textbf{Time on CPU}} & \multicolumn{1}{c}{\multirow{2}{*}{\begin{tabular}[c]{@{}r@{}}\textbf{Space} \\ (GB)\end{tabular}}}\\ \cline{2-9} 
 & \multicolumn{1}{c}{\textbf{Pre.}} & \multicolumn{1}{c}{\textbf{Emb.}} & \multicolumn{1}{c}{\textbf{Cmp.}} & \multicolumn{1}{c|}{\textbf{Total}} & \multicolumn{1}{c}{\textbf{Pre.}} & \multicolumn{1}{c}{\textbf{Emb.}} & \multicolumn{1}{c}{\textbf{Cmp.}} & \multicolumn{1}{c|}{\textbf{Total}} & \\ \hline \hline
DTW & 0.10 & - & 0.58 & 0.68 & 0.01 & - & 28.32 & 28.33 & \textbf{0.02} \\
ERP & 0.10 & - & 0.70 & 0.80 & 0.01 & - & 55.10 & 55.11 & \textbf{0.02} \\
Fr\'echet & 0.10 & - & 0.57 & 0.67 & 0.01 & - & 26.03 & 26.04 & \textbf{0.02} \\
Hausdorff & 0.10 & - & 0.63 & 0.73 & 0.01 & - & 18.71 & 18.72 & \textbf{0.02} \\ \cdashline{1-10} 
NEUTRAJ & 0.49 & 0.46 & 1e-4 & 0.96 & 0.51 & 0.78 & 2e-3 & \textbf{1.29} & 0.13 \\ 
TMN & - & - & - & - & - & - & - & - & - \\
T3S & 0.48 & 0.03 & 1e-4 & 0.51 & 0.49 & 5.09 & 2e-3 & 5.63 & 3.02 \\
TrajCL & 0.47 & 0.02 & 1e-4 & \textbf{0.49} & 0.46 & 1.75 & 2e-3 & 2.21 & 5.81 \\
TrjSR & 0.62 & 0.90 & 1e-4 & 1.52 & 0.63 & 36.89 & 2e-3 & 37.52 & 9.12 \\
TrajGAT & 17.96 & 1.81 & 1e-4 & 19.77 & 18.23 & 17.30 & 2e-3 & 35.53 & 13.89 \\ \hline
{STEDR} (ST) & 0.14 & - & 0.59 & 0.73 & 0.02 & - & 41.85 & 41.87 & \textbf{0.02} \\
{CDDS} (ST) & 0.14 & - & 0.50 & \textbf{0.64} & 0.02 & - & 20.97 & 20.99 & \textbf{0.02} \\ 
\rvcolor{{SAR} (ST)} & \rvcolor{0.21} & \rvcolor{-} & \rvcolor{0.79} & \rvcolor{1.00} & \rvcolor{0.02} & \rvcolor{-} & \rvcolor{24.54} & \rvcolor{24.57} & \rvcolor{0.03} \\ \cdashline{1-10} 
{RSTS} (ST) & 4.98 & 0.08 & 1e-4 & 5.06 & 5.03 & 1.18 & 2e-3 & \textbf{6.30} &  0.76 \\ \hlineB{3}
\end{tabular}
}
\end{table}

\subsection{Trajectory Similarity Queries}\label{subsec:exp:knn}
Next, we report results on trajectory similarity (i.e., $k$NN) queries.
\subsubsection{Setup.}\label{subsubsec:exp:knn_setting}
The default trajectory dataset $D$ contains 100,000 randomly sampled trajectories each with  20 to 200 points. The query set $Q$ contains 1,000 trajectories in the same length range randomly sampled outside $D$.  
The query parameter $k$ is 50 by default. 

We deploy \emph{dedicated indices} for the non-learned and the learned measures, respectively, to provide a fairer comparison. This differs from the existing studies~\cite{neutraj,t2vec} that use R-trees~\cite{r*tree} to index the raw trajectories for query pruning and compute the similarity by the embeddings.
For the non-learned measures, we use two recent indices, i.e., DITA~\cite{dita} for DTW and ERP, and DFT~\cite{DFT} for Hausdorff and Fr\'echet (cf.~Section~\ref{subsec:related_knn}).
For the learned measures, we use generic vector similarity search indices for trajectory embedding-based $k$NN queries, as there are \emph{no existing indices designed specifically} for trajectory embeddings.
We use Faiss~\cite{faiss} which is a generic, widely used similarity search library for vectorized data. The \emph{Inverted file index} \textbf{(IVF)} from this library is used by default for its good balance between query efficiency and accuracy.

We report results on both query efficiency and effectiveness (i.e., accuracy) , as the effectiveness of the learned measures highly impact their applicability. In particular, 
the learned measures are inaccurate. Results on both aspects together guide the choice of learned measures for different application scenarios.
Following the literature~\cite{neutraj,t3s,trajgat}, we use hit ratios (\textbf{HR@50}) to measure the accuracy of the learned measures.
We also report the cost of index construction.
We use all learned measures expect TMN which cannot be applied in index-based $k$NN queries because it does not produce embeddings in advance. Like before, we focus on  batched processing on Xi'an.

\subsubsection{Overall Results.}
We report the elapsed times of index building and query processing in Table~\ref{tab:knn_overview}.
The \emph{index building time} in the table includes the time to build the indices and -- for the learned measures -- the time to compute embeddings for the data trajectories. Likewise, the \emph{query time} for the learned measures includes the time to compute an embedding for the query trajectory (as the query trajectories may arrive online).
Overall, the learned measures take more time on index building, for faster query processing. 

\begin{table}[h]
\small
\centering
\caption{$k$NN index building and query performance results}
\label{tab:knn_overview}
\setlength{\tabcolsep}{0.7mm}
\renewcommand{\arraystretch}{1.1}
\resizebox{0.6\columnwidth}{!}{%
\begin{tabular}{l|rr|r|r}
\hlineB{3}
\multirow{2}{*}{\textbf{Measure}} & \multicolumn{2}{c|}{\textbf{Index building}} & \multicolumn{1}{c|}{\multirow{2}{*}{\begin{tabular}[c]{@{}c@{}} \textbf{Query time} \\ \textbf{(GPU)} \end{tabular}}} & \multicolumn{1}{c}{\multirow{2}{*}{\begin{tabular}[c]{@{}c@{}}\textbf{Query time}\\ \textbf{(CPU)} \end{tabular}}} \\ \cline{2-3} 
 & \multicolumn{1}{c|}{\textbf{Time}} & \multicolumn{1}{c|}{\textbf{Space} (GB)} & \multicolumn{1}{c|}{} & \multicolumn{1}{c}{} \\ \hline \hline
DTW & 0.71 & \textbf{1.60} & 1193.12 & 5441.06  \\
ERP & \textbf{0.69} & \textbf{1.60} & 1304.40 & 6792.25 \\
Fr\'echet & 28.46 & 2.63 & 878.89 & 1783.95 \\
Hausdorff & 28.46 & 2.63 & 903.41 & 3410.28 \\ \cdashline{1-5}
NEUTRAJ & 49.15 & 2.55 & 0.98 & 14.30  \\
T3S & 58.39 & 2.55 & 0.96 & 8.58  \\
TrajCL & 47.88 & 2.55 & \textbf{0.89} & \textbf{4.33} \\
TrjSR & 171.15 & 2.55 & 2.42 & 23.43  \\
TrajGAT & 661.69 & 2.55 & 6.62 & 60.45 \\ 
\hlineB{3}
\end{tabular}
}
\end{table}

\textbf{Index building costs.}
The index build times of the learned measures are higher because they need to first encode raw trajectories into embeddings. In comparison, the non-learned measures index raw trajectory points or segments, and their indices are faster to build. 
The DITA indices are the fastest to build, because they only index a few pivot points of each trajectory. Their space costs (i.e., the index size) are hence also the smallest. DFT indexes  all trajectory segments which has the highest space costs. 

\textbf{$k$NN query costs.}
As for $k$NN querying, the learned measures outperform the non-learned ones significantly, by some two orders of magnitude. 
The fastest learned measure TrajCL is 984 times and 410 times faster than the fastest non-learned measure Fr\'echet when querying on GPU and CPU, respectively. 
Both DFT and DITA are spatial indices, which suffer in their pruning capability when indexing objects with highly skewed aspect ratios, such as trajectories which are long and thin. The trajectories that cannot be pruned require expensive similarity computations with the non-learned measures. 
In contrast, the learned measures use the vector index IVF.
The embeddings are pre-computed, and the similarity scores are now computed by simple embedding scans, which is highly efficient. 
This set of results confirms an important advantage of the learned measures, i.e., their embeddings can be indexed that enable fast $k$NN queries, even using just generic vector indices. Such results motivate further development of dedicated indices for even faster  embedding-based $k$NN query processing.

\begin{table}[h]
\small
\centering
\caption{$k$NN query accuracy (HR@50 of learned measures to approximate non-learned measures)}
\label{tab:effectiveness}
\renewcommand{\arraystretch}{1.1}
\resizebox{0.55\columnwidth}{!}{
\begin{tabular}{l|cccc}
\hlineB{3}
 & \textbf{DTW} & \textbf{ERP} & \textbf{Fr\'echet} & \textbf{Hausdorff} \\ \hline \hline
NEUTRAJ & \textbf{0.629} & 0.417 & 0.671 & 0.678 \\
T3S & 0.514 & \textbf{0.756} & 0.756 & 0.657 \\
TrajCL & 0.528 & 0.421 & \textbf{0.814} & \textbf{0.831} \\
TrjSR & 0.521 & 0.377 & 0.630 & 0.757 \\
TrajGAT & 0.561 & 0.287 & 0.320 & 0.286 \\ 
\hlineB{3}
\end{tabular}
}
\end{table}

\textbf{Overall $k$NN query accuracy.} In Table~\ref{tab:effectiveness}, the HR@50 values indicate how accurately the top 50 trajectories returned by the learned measures approximate those returned by the non-learned measures (recall that the learned measures are trained to approximate the non-learned ones). 
Overall, none of the learned measures return fully accurate results, and their HR@50 are lower than 0.7 in most cases, suggesting opportunities to explore models that can better approximate the non-learned measures. While there are exceptions, the models appear to achieve higher HR@50 scores approximating Fr\'echet and Hausdorff than approximating DTW and ERP. This suggests that DTW and ERP are more difficult to be approximated (by the learned measures tested) than Fr\'echet and Hausdorff -- similar observations have been reported for DTW~\cite{trajgat,t3s}. TrajCL, in particular, achieves HR@50 $> 0.8$ for these two measures.

\begin{figure}[ht]
    \vspace{1mm}
    \includegraphics[width=0.55\textwidth]{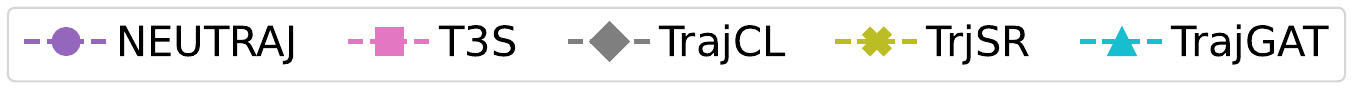} \\
    \subfloat[{DTW}~\label{fig:exp_knn_train_hr50_dtw}]{
        \includegraphics[width=0.33\textwidth]{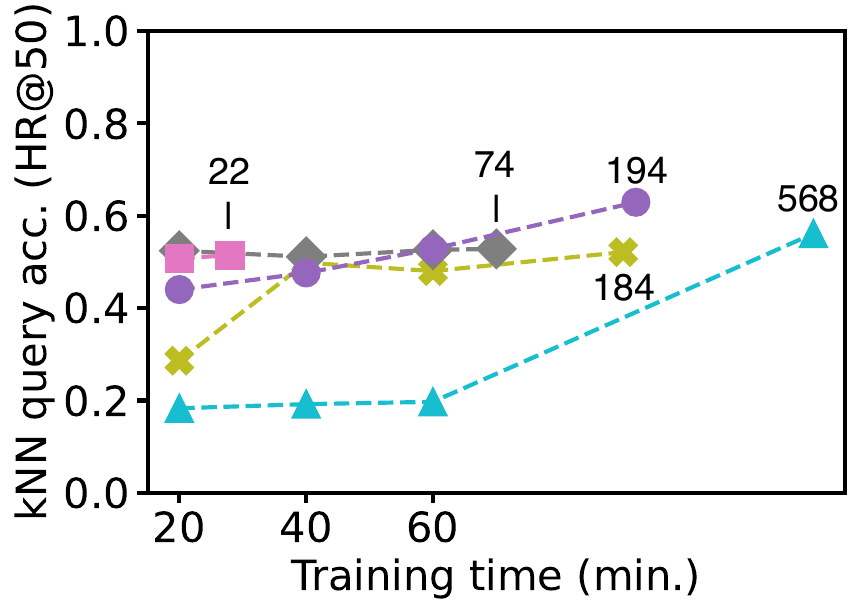}
    }\hspace{1mm}
    \subfloat[{ERP}~\label{fig:exp_knn_train_hr50_erp}]{
        \includegraphics[width=0.33\textwidth]{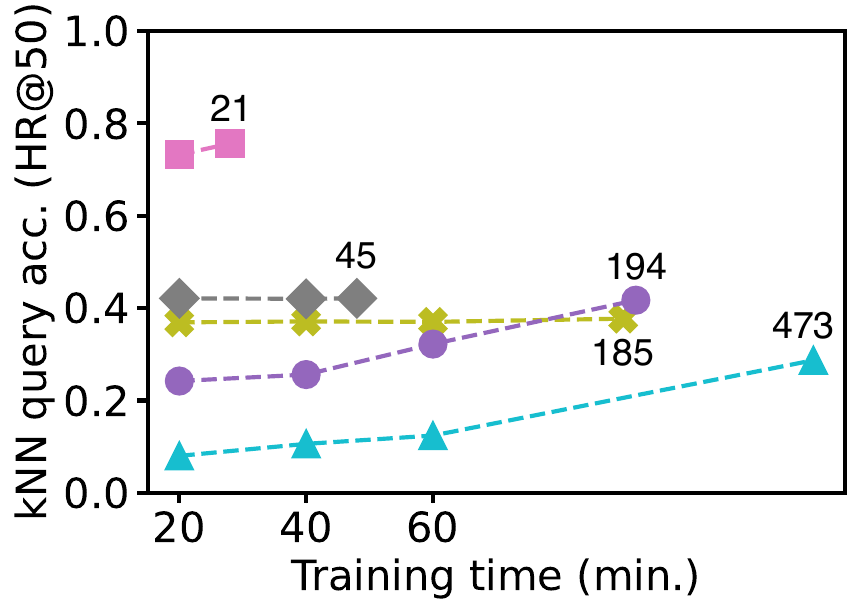}
    } \\
    \subfloat[{Fr\'echet}~\label{fig:exp_knn_train_hr50_frechet}]{
        \includegraphics[width=0.33\textwidth]{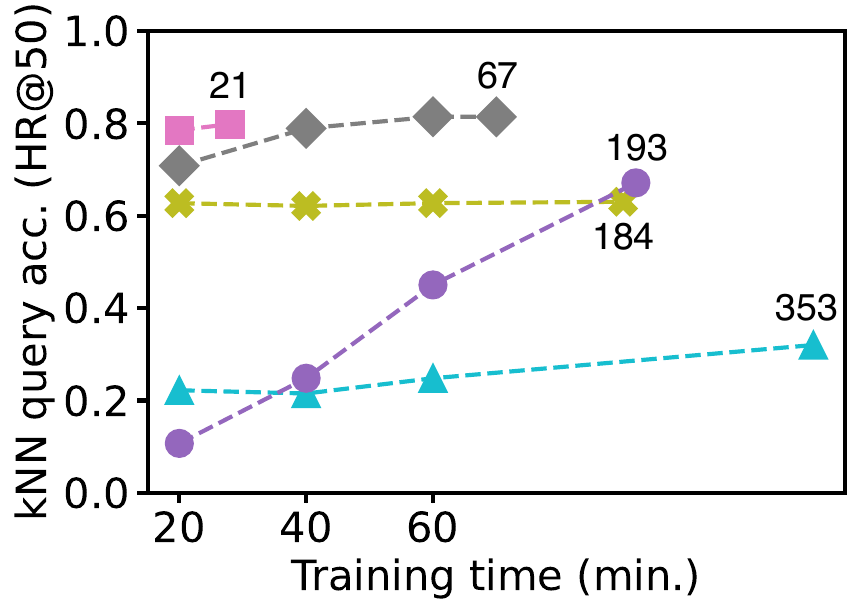}
    }\hspace{1mm}
    \subfloat[{Hausdorff}~\label{fig:exp_knn_train_hr50_hausdorff}]{
        \includegraphics[width=0.33\textwidth]{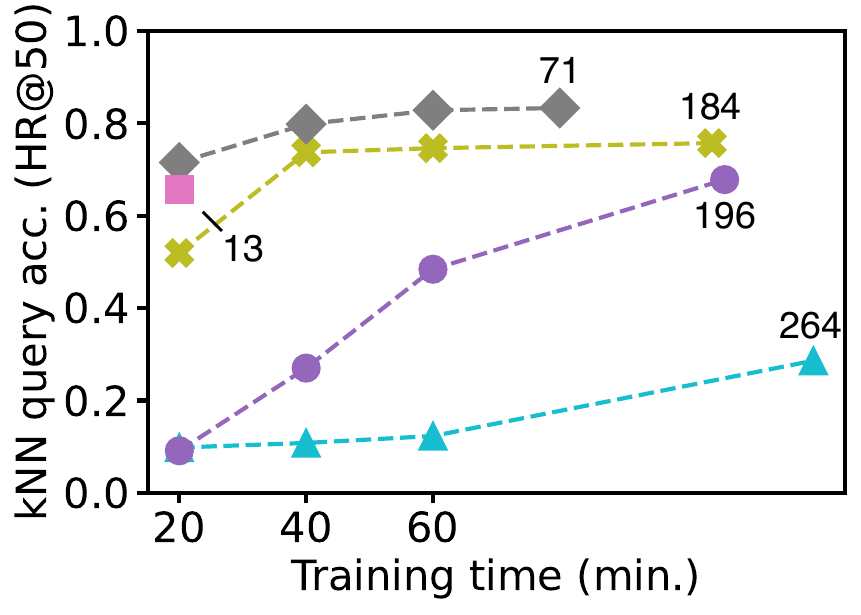}
    }
    \vspace{1mm}
    \caption{Training time of the learned measures vs. $k$NN query accuracy (the numbers are model convergence times)}\label{fig:exp_knn_train_hr50}
\end{figure}

\textbf{$k$NN query accuracy vs. trajectory similarity model training time.} 
Figure~\ref{fig:exp_knn_train_hr50} reports HR@50 of the learned measures, where the model training time is constrained from 20 to 60 minutes. Note that the trajectory similarity model training time is excluded from the index build times reported above (and is excluded from all the other sets of results), i.e., the last set of experiments assumes trained similarity measures. This set of results further provides guidance on model selection for applications with time constraints on system preparation from scratch or model retraining. 

The figure shows that
when the model training time is very limited, e.g., 20 minutes, T3S is generally the best option, except that TrajCL is 6\% more accurate than T3S when approximating Hausdorff.
When the training time increases to 40 minutes, TrajCL performs the best to approximate the non-learned measures except for the ERP (for which T3S remains the best). 
This is because T3S uses a vanilla attention model that is faster to train, while TrajCL uses a multi-view attention model that takes more time to train but may produce higher accuracy given enough time. 
\emph{These results also confirm the superiority of using attention models to learn trajectory similarity, especially when the training time is limited.}

\subsubsection{Impact of Indices for Learned Measures.}\label{subsubsec:exp:knn_indices}
We compare the default IVF index with the flat index (\textbf{FLAT}) and the hierarchical navigable small world index (\textbf{HNSW})~\cite{hnsw} for trajectory $k$NN queries. FLAT stores the full trajectory embeddings of the input dataset $D$ and performs full scans for similarity searches. HNSW creates a navigable small world graph based on the embeddings and uses hierarchical skip lists to link graph nodes to speed up the search. Both HNSW and IVF return approximate query results.
We repeat the experiments of batched $k$NN queries on Xi'an. We report the index build time, the query time on GPU, and the query accuracy to approximate the Hausdorff measure.

Table~\ref{tab:knn_index} presents the results. Overall, the three indices have similar build times, which include and are dominated by the embedding computation times. HNSW takes the longest to build because of its complex graph structure, which in turn helps it to achieve the fastest query times while sacrificing the accuracy. On the opposite, FLAT is the fastest to build, while it is the slowest for query processing because of its brute-force query strategy. FLAT is still inaccurate, because the embeddings are approximations of the raw trajectories. IVF has the best balanced performance. It builds as fast as FLAT, is fast at query processing, and has a good accuracy.

\begin{table}[ht]
\centering
\caption{Index comparison for the learned measures}
\label{tab:knn_index}
\setlength{\tabcolsep}{0.8mm}
\renewcommand{\arraystretch}{1.1}
\resizebox{0.7\columnwidth}{!}{%
\begin{tabular}{l|rrr|rcr|rcr}
\hlineB{3}
\multirow{2}{*}{\textbf{Measure}} & \multicolumn{3}{c|}{\textbf{Building time}} & \multicolumn{3}{c|}{\textbf{Query time}} & \multicolumn{3}{c}{\textbf{Query accuracy}} \\ \cline{2-10} 
 & \multicolumn{1}{c}{\textbf{FLAT}} & \multicolumn{1}{c}{\textbf{HNSW}} & \multicolumn{1}{c|}{\textbf{IVF}} & \multicolumn{1}{c}{\textbf{FLAT}} & \multicolumn{1}{c}{\textbf{HNSW}} & \multicolumn{1}{c|}{\textbf{IVF}} & \multicolumn{1}{c}{\textbf{FLAT}} & \multicolumn{1}{c}{\textbf{HNSW}} & \multicolumn{1}{c}{\textbf{IVF}} \\ \hline \hline
NEUTRAJ & 48.87 & 53.22 & 49.15 & 4.09 & 0.61 & 0.98 & 0.717 & 0.661 & 0.678 \\
T3S & 57.69 & 69.99 & 58.39 & 3.91 & 0.51 & 0.96 & 0.694 & 0.628 & 0.657 \\
TrajCL & 46.92 & 58.31 & 47.88 & 3.90 & 0.43 & 0.89 & 0.879 & 0.819 & 0.831 \\
TrjSR & 162.63 & 183.42 & 171.15 & 6.32 & 1.33 & 2.42 & 0.796 & 0.731 & 0.757 \\
TrajGAT & 660.74 & 667.25 & 661.69 & 10.13 & 6.19 & 6.62 & 0.327 & 0.259 & 0.286 \\ \hlineB{3}
\end{tabular}%
}
\end{table}


\subsubsection{Impact of the Size of the Trajectory Dataset $|D|$}

\begin{figure}[ht]
    \hspace{5mm}
    \includegraphics[width=0.55\textwidth]{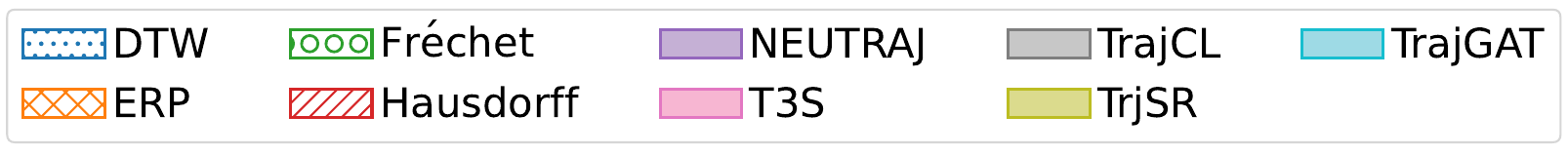} \\
    \vspace{-2mm}
    \subfloat[{Index build time}~\label{fig:exp_knn_numtrajs_gpu_build}]{
        \includegraphics[width=0.33\textwidth]{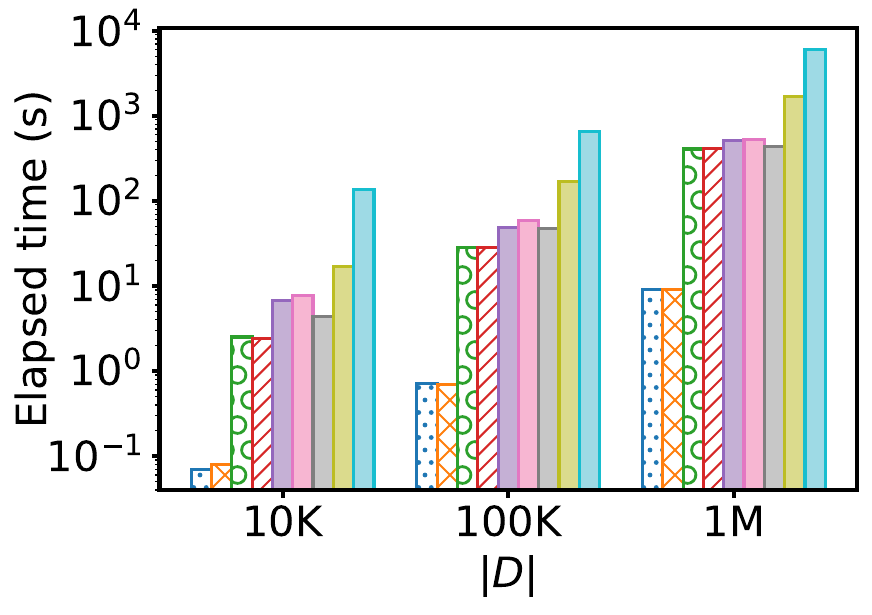}
    }
    \subfloat[{$k$NN query time}~\label{fig:exp_knn_numtrajs_gpu_query}]{
        \includegraphics[width=0.33\textwidth]{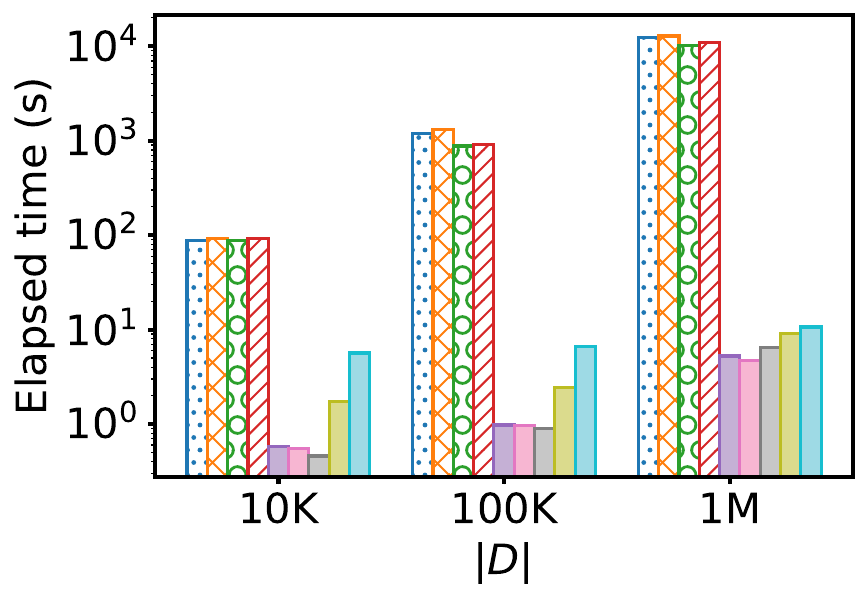}
    }\vspace{1mm}
    \caption{$k$NN index building and query time vs. the size of the trajectory dataset}\label{fig:exp_knn_numtrajs_gpu}
\end{figure}

We vary the size of the trajectory dataset $|D|$ from $10^4$ to $10^6$. Here, we only report the results on GPU, as similar patterns are observed on CPU.
As Figure~\ref{fig:exp_knn_numtrajs_gpu_build} shows, the index build times of all measures grow roughly linearly with $|D|$. The indices of the non-learned measures are consistently faster to build, by up to three orders of magnitude (0.074 vs. 134.996 on DTW and TrajGAT when $|D| = 10^4$).

Figure~\ref{fig:exp_knn_numtrajs_gpu_query} shows the $k$NN query times. Like in  Table~\ref{tab:knn_overview},  the non-learned measures are one to three orders of magnitude slower than the learned ones. Importantly, as $|D|$ increases, the performance gap increases, attributing to the simple searches of trajectory embeddings on IVF for the learned measures. 
\emph{Since there are no universally applicable trajectory similarity measures, unless accurate results based on a specific non-learned measure are required, the learned measures are the more practical choice for $k$NN queries.} 

\subsubsection{Impact of the Number of Points in the Query  Trajectories.}
We vary the number of points in the query trajectories from 20 to 1,600. 
The comparative performance between the non-learned and learned measures again resembles those reported earlier. We omit the result figure for conciseness.  

\subsection{Trajectory Clustering}\label{subsec:exp:clustering}
Next, we study the efficiency and effectiveness of trajectory clustering.

\subsubsection{Setup}
We follow the commonly used trajectory clustering paradigm~\cite{trajclustering_yaodi, detect}.
For the non-learned measures, we apply the $k$-medoids algorithm~\cite{kmedoids} to cluster the raw trajectories.
For the learned measures, we apply $k$-medoids on the  embeddings.

The default dataset $D$ consists of 1,000 trajectories that are randomly sampled from a dataset, where the number of points on trajectories $n$ is between 20 and 200. The number of clusters is set to 10 by default (i.e., $k=10$ in $k$-medoids). 

Trajectory clustering based on learned measures is also an approximation of that based on non-learned measures.
We follow the literature and use the \emph{rand index} (RI) to evaluate clustering accuracy. RI computes the percentage of the ground-truth similar trajectory pairs (derived based on non-learned measures) that are assigned to the same cluster. We again report results on  Xi'an. 

\begin{table}[htp]
\centering
\caption{Detailed time and space costs of batched trajectory clustering on Xi'an}
\label{tab:clustering_time}
\setlength{\tabcolsep}{0.5mm}
\renewcommand{\arraystretch}{1.1}
\resizebox{0.7\columnwidth}{!}{
\begin{tabular}{l|rrrr|rrrr|r}
\hlineB{3} 
\multirow{2}{*}{\textbf{Measure}} & \multicolumn{4}{c|}{\textbf{Time on GPU}} & \multicolumn{4}{c|}{\textbf{Time on CPU}} & \multicolumn{1}{c}{\multirow{2}{*}{\begin{tabular}[c]{@{}r@{}}\textbf{Space} \\ (GB)\end{tabular}}}\\ \cline{2-9} 
 & \multicolumn{1}{c}{\textbf{Pre.}} & \multicolumn{1}{c}{\textbf{Emb.}} & \multicolumn{1}{c}{\textbf{Clst.}} & \multicolumn{1}{c|}{\textbf{Total}} & \multicolumn{1}{c}{\textbf{Pre.}} & \multicolumn{1}{c}{\textbf{Emb.}} & \multicolumn{1}{c}{\textbf{Clst.}} & \multicolumn{1}{c|}{\textbf{Total}} & \\ \hline \hline
DTW & 0.60 & - & 3.22 & 3.82 & 0.24 & - & 226.09 & 226.33 & \textbf{0.02} \\
ERP & 0.60 & - & 3.70 & 4.29 & 0.25 & - & 436.48 & 436.73 & \textbf{0.02} \\
Fr\'echet & 0.60 & - & 3.26 & 3.86 & 0.24 & - & 223.68 & 223.92  & \textbf{0.02} \\
Hausdorff & 0.61 & - & 3.37 & 3.98 & 0.25 & - & 167.60 & 167.85  & \textbf{0.02} \\ \cdashline{1-10} 
NEUTRAJ & 0.26 & 0.17 & 0.27 & \textbf{0.70} & 0.35 & 0.87 & 0.21 & \textbf{1.43} & 0.88  \\ 
T3S & 0.40 & 0.02 & 0.31 & 0.74 & 0.47 & 2.26 & 0.34 & 3.06 & 2.74 \\
TrajCL & 0.36 & 0.01 & 0.33 & \textbf{0.70} & 0.40 & 1.40 & 0.31 & 2.11 & 2.35 \\
TrjSR & 0.52 & 0.05 & 0.41 & 0.97 & 0.58 & 17.12 & 0.31 & 18.01 & 9.12 \\
TrajGAT & 6.82 & 0.35 & 0.22 & 7.39 & 7.56 & 6.87 & 0.32 & 14.75 & 2.08 \\ \hline

\rvcolor{STEDR} & \rvcolor{0.79} & \rvcolor{-} & \rvcolor{3.64} & \rvcolor{4.42} & \rvcolor{1.09} & \rvcolor{-} & \rvcolor{386.26} & \rvcolor{387.34} & \rvcolor{\textbf{0.02}} \\
\rvcolor{CDDS} & \rvcolor{0.82} & \rvcolor{-} & \rvcolor{3.35} & \rvcolor{\textbf{4.17}} & \rvcolor{1.08} & \rvcolor{-} & \rvcolor{171.42} & \rvcolor{172.50} & \rvcolor{\textbf{0.02}} \\
\rvcolor{SAR} & \rvcolor{0.81} & \rvcolor{-} & \rvcolor{3.92} & \rvcolor{4.73} & \rvcolor{1.09} & \rvcolor{-} & \rvcolor{167.04} & \rvcolor{168.13} & \rvcolor{0.03} \\ \cdashline{1-10} 
\rvcolor{RSTS} & \rvcolor{6.22} & \rvcolor{0.05} & \rvcolor{0.31} & \rvcolor{6.58} & \rvcolor{6.37} & \rvcolor{0.97} & \rvcolor{0.31} & \rvcolor{\textbf{7.65}} & \rvcolor{0.38} \\ 
\hlineB{3}
\end{tabular}
}
\end{table}

\subsubsection{Overall Results.} 
Tables~\ref{tab:clustering_time} and~\ref{tab:clustering_effectiveness} report the results.  
\emph{Overall, the learned measures are faster than the non-learned ones, while they take more memory space and serve inaccurate results.}

\textbf{Clustering costs.} 
We use the clustering time (denoted as \textbf{``Clst.''}) to measure the time to cluster the raw trajectories with the non-learned measures or to cluster the embeddings for the learned measures. We only compute the embeddings once, which are reused through out the clustering process of a dataset. Thus, the learned measures offer faster clustering (while they require higher space costs to store the embeddings), as shown in Table~\ref{tab:clustering_time}.
For the same reason, the embedding times (\textbf{``Emb.''}) reported in Table~\ref{tab:clustering_time} are much lower than those reported in Table~\ref{tab:comp_detail_time_pairs} where the embeddings are recomputed every time when a learned measure is computed.

\begin{table}[h]
\small
\centering
\caption{Clustering accuracy (RI of learned measures to approximate non-learned measures)}
\label{tab:clustering_effectiveness}
\renewcommand{\arraystretch}{1.1}
\resizebox{0.55\columnwidth}{!}{
\begin{tabular}{l|cccc}
\hlineB{3}
 & \textbf{DTW} & \textbf{ERP} & \textbf{Fr\'echet} & \textbf{Hausdorff} \\ \hline \hline
NEUTRAJ & \textbf{0.885} & 0.811 & 0.811 & \textbf{0.874} \\
T3S & 0.832 & 0.816 & 0.823 & 0.820 \\
TrajCL & 0.816 & \textbf{0.817} & \textbf{0.827} & 0.811 \\
TrjSR & 0.708 & 0.718 & 0.746 & 0.799 \\
TrajGAT & 0.808 & 0.646 & 0.712 & 0.433 \\ 
\hlineB{3}
\end{tabular}
}
\end{table}

\textbf{Clustering accuracy.}
Table~\ref{tab:clustering_effectiveness} shows the  accuracy results.  
Similar to the $k$NN results in Table~\ref{tab:effectiveness}, NEUTRAJ is strong for DTW, while TrajCL approximates ERP and Fr\'echet well, and no learned measures can approximate the non-learned ones fully accurately.


\subsubsection{Impact of the Size of Trajectory Dataset $|D|$} 
\begin{figure}[ht] 
    \includegraphics[width=0.7\textwidth]{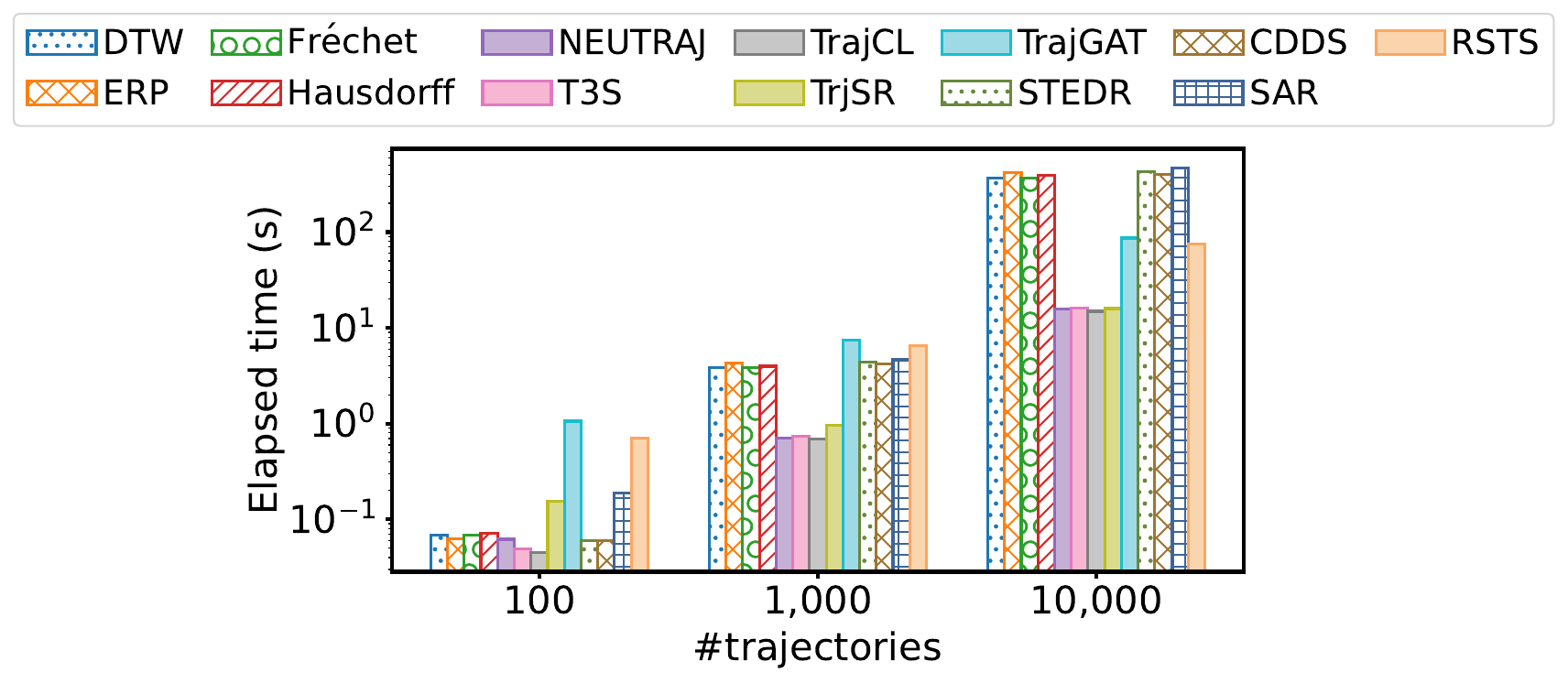}
    \caption{Clustering time vs. the number of trajectories}\label{fig:exp_clustering_numtrajs_gpu}
\end{figure}

We vary the size of the trajectory dataset $|D|$ from 100 to 10,000. Figure~\ref{fig:exp_clustering_numtrajs_gpu} shows the time of batch clustering on GPU. With $|D|$  increasing, clustering takes more time with both types of measures. Clustering with non-learned measures generally consumes more time than that with the learned ones. 
When datasets are small (i.e., $|D|=100$ for the spatial measures or $|D|\leqslant1,000$ for the spatio-temporal measures), the non-learned measures can be faster, because the one-off embedding computation times dominate in this setting.

We also varied the number of points in trajectories. Since the comparative patterns have not changed, we omit the result figure.

\section{Future Work}\label{subsec:exp:discussion}
The empirical study reveals that: 
(1)~The learned measures are slower than the non-learned ones when computing trajectory similarity on the fly. 
(2)~The learned measures are faster than the non-learned ones for $k$NN queries and clustering when the embeddings can be pre-computed, although there is no accuracy guarantee. 
(3)~Self-attention network-based measures are relatively fast to train to obtain a high accuracy of trajectory similarity.

These results motivate important questions for future work: 

RQ1: Can learned measures also outperform the non-learned ones for online similarity computation? 

RQ2: Can a learned measure approximate a wide range of different non-learned measures all with a high accuracy?

RQ3: Can learned measures approximate the non-learned ones with accuracy guarantees? 

RQ4: Can non-learned measures be indexed to achieve higher $k$NN query efficiency?

Here, we present an attempt to answer the first two questions. 

To answer RQ1, we make use a \emph{feedforward neural network} (FFN) as the trajectory encoder, which is arguably the simplest and most  efficient deep learning model. 
We use a two-layer FFN, where the size of the intermediate vectors is $d$ and the activation function is \emph{ReLU}. The inputs are raw two-dimensional coordinates of the points on a trajectory, for efficiency considerations. 

\begin{table}[h]
\centering
\caption{Comparison between the existing representative measures and an FFN-based measure}
\label{tab:exp_discussion_mlp}
\renewcommand{\arraystretch}{1.05}
\resizebox{0.7\columnwidth}{!}{
\begin{tabular}{l|c|cc|cc}
\hlineB{3}
 & \multicolumn{1}{c|}{\textbf{Similarity comp.}} & \multicolumn{2}{c|}{\textbf{$k$NN query}} & \multicolumn{2}{c}{\textbf{Clustering}} \\ \cline{2-6} 
 & \multicolumn{1}{c|}{\textbf{Time (s)}} & \multicolumn{1}{c|}{\textbf{Time (s)}} & \multicolumn{1}{c|}{\textbf{ HR@50}} &\multicolumn{1}{c|}{\textbf{Time (s)}} & \multicolumn{1}{c}{\textbf{ RI}} \\\hline \hline
Hausdorff & \textbf{5.97} & 903.41 & \textbf{100.0\%} & 3.98 & \textbf{100.0\%}\\ \cdashline{1-6}
TrajCL & 77.40 & 0.89 & 83.1\% & 0.70 & 81.1\% \\
\textbf{FFN} & 6.20 & \textbf{0.41} & 59.1\% & \textbf{0.41} & 50.2\% \\ 
\hlineB{3}
\end{tabular}
}
\end{table}

We follow the previous default experimental settings of batched computation on GPU and repeat the experiments with FFN to approximate Hausdorff. We compare the results with the most efficient non-learned and learned measures, i.e., Hausdorff and TrajCL, respectively.
As Table~\ref{tab:exp_discussion_mlp} shows, FFN is much more efficient than TrajCL for trajectory similarity computation, and it even achieves a comparable time to that of Hausdorff. 
Such results are achieved because the time complexity of FFN for trajectory similarity computation is $O(nd)$, which is lower than that of TrajCL and similar to that of Hausdorff.
However, this high efficiency comes with a substantial cost in the $k$NN query and clustering accuracy. 

Such results show the potential of learned measures to obtain a high efficiency for trajectory similarity computation, thus meeting their original promise, while they also highlight the challenges in obtaining high query and clustering accuracy at the same time. 

For RQ2, we argue that \emph{large language models} (LLM)~\cite{gpt1,llama1} have a strong potential to learn a generic measure for approximating different non-learned measures. LLMs are trained on large text corpora to learn sequential patterns of texts. Since trajectories are also sequences, it would be interesting to study how LLMs can be adapted for numeric sequences. 
When an LLM is trained with a large trajectory corpus, it could be instruction-tuned to compute different trajectory similarity measures given different input prompts.

In addition, our study has focused on trajectory similarity in a low-dimensional Euclidean space. There are rich studies on trajectories over road networks~\cite{gts,st2vec,tsjoin,torch} as well as multi-dimensional time series (high dimensional trajectories)~\cite{multitimesimi_1,multitimesimi_2,multitimesimi_3,multitimesimi_4}. An empirical study with such data will also be an interesting future work.

\section{Conclusion}\label{sec:conclusion}
We revisited both non-learned and deep learning-based trajectory similarity measures and studied their empirical efficiency comprehensively. We found that the learned measures outperform the non-learned measures  as promised in literature, only when the trajectory embeddings can be pre-computed.
Meanwhile, such measures lack accuracy when approximating the non-learned measures. Among the learned measures, the self-attention-based ones are the fastest to train and also offer the highest accuracy. 
In comparison, the non-learned measures do not require pre-computation and are more suitable for one-off trajectory similarity computation.
These results open up research opportunities in designing advanced learned measures with even higher efficiency and accuracy.

\begin{acks}
This work is partially supported by Australian Research Council (ARC) Discovery Projects DP230101534 and DP240101006. Gao Cong is supported in part by a Singapore MOE AcRF Tier-2 grant MOE-T2EP20221-0015 and a Singapore MOE AcRF Tier-1 project RT6/23.
\end{acks}

\balance
\bibliographystyle{ACM-Reference-Format}
\bibliography{main}

\newpage

\appendix

\section{GPU-based Non-Learned Measure Implementation}\label{sec:implementation}
We present three \emph{meta-algorithms} for GPU-based implementations of the three types of non-learned trajectory similarity measures as described in Section~\ref{subsec:related_heuristic}, to enable empirical comparisons with the learned measures on GPUs.

\subsection{Parallelizing Linear Scan-Based Measures}
\begin{figure}[htp]
   \centering
   \includegraphics[width=0.33\textwidth]{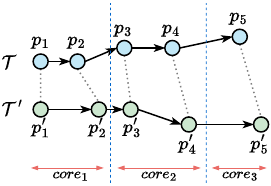}
   \caption{Parallelization of linear scan-based measures}
   \label{fig:linear_core}
\end{figure}

Linear scan-based non-learned measures first pair up points from two trajectories sequentially, e.g., by  point indices (in SPD) or timestamps (in CDDS and SAX). Then, they compute a similarity score (distance) for each point pair and aggregate the scores to obtain the overall trajectory similarity score. As the similarity scores of the point pairs are independent from each other, we can simply partition each trajectory into sub-trajectories and parallelize the processing of the sub-trajectories. We name such an algorithm \textbf{\texttt{par-scan}} and illustrate it with Figure~\ref{fig:linear_core}, where the two input trajectories $\mathcal{T}$ and $\mathcal{T}'$ are divided into three pairs of sub-trajectories sequentially. 
We distribute each pair onto a computation core (i.e., a GPU core, same below), where $core_i$ denotes the $i$-th core, and each pair is processed  concurrently. 
Each core computes the distances for at most $\max(|\mathcal{T}|, |\mathcal{T}'|)/n_c$ point pairs, assuming $n_c$ cores on the GPU. 
In the end, the result from each core is aggregated sequentially. 
We omit the pseudo-code as it is straightforward. 

\subsection{Parallelizing DP-Based Measures}
\begin{figure}[h!]
    \centering
    \begin{minipage}[b]{0.28\textwidth}
        \subfloat[{ }~\label{fig:dp_gpu_1}]{
            \includegraphics[width=0.965\textwidth]{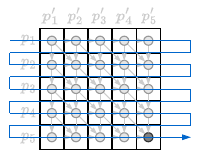}
        } \\ 
        \subfloat[{ }~\label{fig:dp_gpu_2}]{
            \includegraphics[width=\textwidth]{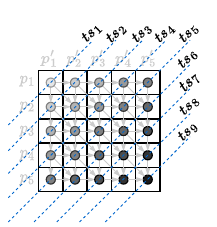}
        }
    \end{minipage}
    \hspace{1.5mm}
    \begin{minipage}[b]{0.30\textwidth}
        \subfloat[{ }~\label{fig:dp_gpu_3}]{
            \includegraphics[width=\textwidth]{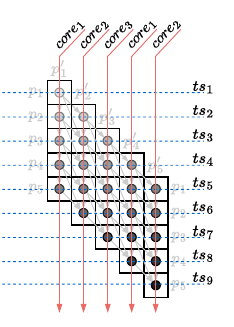} 
        }
        \vspace{3mm}
    \end{minipage}
    \caption{Score matrix computation in DP-based measures: (a)~single core strategy, (b)~anti-diagonal computational dependencies, (c)~parallelized strategy.}\label{fig:dp_gpu} 
\end{figure}

Dynamic programming (DP)-based  measures need to compute an optimal  matching of the points from two trajectories and their similarity scores by filling up a DP \emph{score matrix}. In general, the score matrix is computed row by row, and from the left to the right in each row. The last computed score in the matrix is returned as the final similarity score. 
Figure~\ref{fig:dp_gpu_1} illustrates the process, assuming two trajectories of five points each, i.e., $p_1$ to $p_5$ and $p_1'$ to $p_5'$, respectively. A dot at row $i$, column $j$ represents the intermediate similarity score between two partial trajectories $[p_1, p_2, \ldots, p_i]$ and $[p'_1, p'_2, \ldots, p'_j]$ (the actual similarity value has been omitted for simplicity). The blue arrow denotes the order of computation, and the bottom right dot represents the final similarity score between the two trajectories.  Note the gray arrows between the dots. They show the sequential computational dependency between the intermediate similarity scores, which are the key challenge in parallelization.

\textbf{The parallelization strategy.}
To break the sequential computational dependency of the score matrix computation, we adapt the \emph{anti-diagonal wavefront parallelization} strategy~\cite{dp_parallel1,dp_parallel2,dp_parallel3,dp_parallel4,dp_parallel6}. The core idea is to isolate the elements in the score matrix that are independent from the others, which enable parallel processing.

We use Figure~\ref{fig:dp_gpu_2}
to illustrate the parallelization strategy. It is intuitive 
that the intermediate scores on the  same blue dash line (an anti-diagonal, e.g., $ts_2$) are independent from each other, while they all depend on the scores to their upper left. Thus, the score matrix can be computed one anti-diagonal at a time, from the top left to the bottom right, i.e., from anti-diagonal $ts_1$ to anti-diagonal~$ts_9$.

Figure~\ref{fig:dp_gpu_3} re-plots the score matrix following the processing order of the anti-diagonals strategy. The dots and the gray arrows still represent the scores and their computational dependencies, respectively. Only the positions of the intermediate scores have been reorganized such that their independent relationships become more obvious. Each column in this figure (which corresponds to partial trajectory $[p'_1, p'_2, \ldots, p'_j]$) can be computed by a different computation core in parallel, as long as the core for the $j$-th column starts one time step earlier than that for the $(j+1)$-th column. 

When there are more cores than  points on $\mathcal{T}'$, each point (i.e., a partial trajectory up to the point) is simply assigned to a different core. Otherwise, we take an interleaving strategy and assign one point on $\mathcal{T}'$ to a core sequentially until all cores have been assigned. We then repeat this process from the first core again, as the figure shows (there are three cores). This strategy is better than assigning a continuous segment of $\mathcal{T}'$ to a core, as it reduces the wait time for the cores assigned with the later points on $\mathcal{T}'$.

\textbf{The parallel algorithm.} 
We name such an parallel algorithm based on the anti-diagonals strategy  \textbf{\texttt{par-DP}} and summarize it in Algorithm~\ref{algo:gpu_dp}. 
Since each row of the DP score matrix in Figure~\ref{fig:dp_gpu_3} depends on the two rows above, we create a two-dimensional matrix of three rows and $|\mathcal{T}'|+1$ columns for the computation, denoted by $M$ (Line~2). Here, an extra column has been added for ease of iterative computation.  Note that sequential DP implementations only require two rows, as one of the read rows can be reused for writing. In parallel implementation, three rows are needed to avoid reading dirty data when multiple cores read and write the matrix concurrently. 
Matrix $M$ is initialized as per the requirement of a target  non-learned measure (Line~3), e.g., zeroing $M$ for EDR.

\begin{algorithm}[htp!]
\SetNoFillComment
\caption{Parallel DP-based measure  (\texttt{par-DP})} \label{algo:gpu_dp} 

\KwIn{$\mathcal{T}$ and $\mathcal{T}'$: two trajectories;  
$n_{c}$: the number of cores.
} 
\KwOut{trajectory similarity score}
\BlankLine
$n_1 \leftarrow |\mathcal{T}|$, $n_2 \leftarrow |\mathcal{T}'|$\;
$M \leftarrow array([3, n_2 + 1])$  \tcp*[r]{DP score matrix} 
Initialize $M$ as required by the target non-learned measure\;
$n_{ts} \leftarrow n_1 + n_2 - 1$ \tcp*[r]{No. of time slots (Figure~\ref{fig:dp_gpu_3})}
\For {$ts \in [0, n_{ts})$} { 
    $r\_row_1 \leftarrow ts \% 3$,     $r\_row_2 \leftarrow (ts+1) \% 3$\;
    $w\_row \leftarrow (ts+2) \% 3$\;
    \tcc{The following $while$ block is executed on $n_c$ cores in parallel, where $cid \in [0, n_c)$ is the sequence number of the current core.}

    \While{$cid \in [0, n_2)$} {
        \If {$cid \leqslant ts < cid + n_1$} {
            Compute $subcost$ based on $M[r\_row_1, cid]$, $M[r\_row_2, cid]$, and $M[r\_row_2, cid+1]$  following rules of the target non-learned measure\; 
            $M[w\_row, cid+1] \leftarrow subcost$\; 
        }
        $cid \leftarrow cid + n_c$\;
    }
    Synchronize\;
}
\Return $M[(n_{ts}+1)\%3, n_2]$\;
\end{algorithm}

The score matrix is computed in $n_{ts} = |\mathcal{T}| + |\mathcal{T}'| + 1$ time slots (for the  $n_{ts}$ anti-diagonals, cf. Figure~\ref{fig:dp_gpu_3}). At time slot $ts$, scores in 
the $(ts \% 3)$-th and $[(ts+1) \% 3]$-th rows of matrix $M$ are read, which are used to compute the scores for the next anti-diagonal, to be stored in the $w\_row$-th row (Lines~5 to~7). For each time slot, $n_c$ cores compute in parallel for the up to $|\mathcal{T}'|$ values of an anti-diagonal. The $i$-th value is computed by the ($i \% n_c$)-th core based on the $i$-th value in the second read row and the $(i-1)$-th values of both read rows (Lines~8 to~12). 
The exact computation depends on the target non-learned measure (e.g., maximization for LCSS) and is not our focus. After the computation, we need a synchronization step to ensure that all cores have completed their computations (Line~13). When all $ts$ time slots are computed, the element at the last column of the last write row is the target similarity score (Line~14).

\begin{figure}[htp!]
   \centering
   \includegraphics[width=0.32\textwidth]{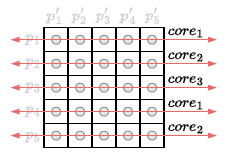}
   \caption{Parallelization of enumeration-based measures}
   \label{fig:enum_core}
\end{figure}

\subsection{Parallelizing Enumeration-Based Measures}

Enumeration-based measures check all combinations of point pairs. Like in the linear scan-based measures, there is no computational dependency between the similarity scores of any two pairs of points. The computation of the scores is thus embarrassingly parallel. We name the parallel algorithm \textbf{\texttt{par-enum}} and illustrate it with Figure~\ref{fig:enum_core}. The pseudo-code is omitted as it is straightforward. 
As the figure shows, we distribute the computation of the similarity scores onto $n_c = 3$ cores where each core computes the scores for $|\mathcal{T}'|/n_c$ points on $\mathcal{T}'$. Here, the interleaving strategy is used again, while parallizing by continuous segments will work just the same.  All similarity cores obtained are harvested to compute the final trajectory similarity on a single core (e.g., the CPU) following the definition of a target enumeration-based measure.

\end{document}